\documentclass[preprint,aps,amsmath,nofootinbib,tightenlines,floatfix]{revtex4}
\usepackage{amssymb}
\usepackage{graphicx,epsfig,amssymb}
\usepackage{multirow}
\usepackage{subfigure}
\usepackage{enumerate}

\def\met{\ensuremath{\not\!\!{E_{T}}}}
\newcommand{\gev}{{\rm GeV}}

\def\beq{\begin{equation}}
\def\eeq{\end{equation}}
\def\bea{\begin{eqnarray}}
\def\eea{\end{eqnarray}}
\def\d{\displaystyle}

\def\tev{\rm TeV}
\def\mhpp{m_{H^{\pm\pm}}}
\def\hpp{H^{\pm\pm}}
\def\d0{D\O\ }

\begin{document}

%\title{Reexamining constraints and discovery reach for same-sign dilepton resonances
%}
\title{A Simplified Model Approach to Same-sign Dilepton Resonances
}
\author{
         Vikram Rentala$^{1,2}$\footnote{vrentala@gmail.com}, 
         William Shepherd$^{1}$\footnote{will.shepherd@gmail.com}, \,and\,
        Shufang Su$^{1,2}$\footnote{shufang@physics.arizona.edu}}
%}
\affiliation{
$^{1}$ Department of Physics, University of Arizona, Tucson, Arizona  85721\\
$^{2}$  Department of Physics and Astronomy, University of California, Irvine, California 92697
}

\begin{abstract}

We discuss  same-sign  dilepton resonances in the simplified model approach. The 
relevant ${\rm SU(3)}_Q^J$ quantum numbers are ${\bf 1}_2^{0,1,2}$. For 
simplicity, we only consider a spin 0 scalar, which is typically referred 
to as a doubly charged Higgs   in the literature. We consider the
three simplest cases where the doubly charged Higgs resides in a singlet, doublet or 
triplet ${\rm SU(2)}_L$ representation. We discuss production and decay of such a
doubly charged Higgs, summarize the current direct search limits, and obtain   mass limits in the cases of singlet and doublet for the first time.  We also  present
a complete set of updated indirect search limits.  We study the discovery potential at the Large Hadron Collider (LHC)
with center of mass energies 7 and 14 TeV for the dominant Drell-Yan pair production with $\hpp$ decay in the 
$ee$ and $\mu\mu$ channels.  We find that at 7 TeV, the LHC with 10 ${\rm fb}^{-1}$ luminosity can probe mass of the doubly charged Higgs up to 380 GeV assuming 100\% decay to leptons.   At 14 TeV, the  LHC with 100 ${\rm fb}^{-1}$ luminosity can reach a  mass of up to 800 GeV.
\end{abstract}

%\pacs{} 

\maketitle

\section{Introduction}
%\vspace{0.5in}
The Large Hadron Collider (LHC) with center of mass energies 7 $-$ 14 TeV is  probing the energy frontier of particle physics at the Electroweak scale.    While the existing experimental searches at the LHC will play an important role in constraining     new physics models, or to discovering new physics signatures, most of the current analyses are done under a particular model framework or a specific model parameterizations.  The advantage of performing a model specific study is that cuts and analyses can be optimized for that particular model or model parameterization to maximize the reach and sensitivity.  The disadvantage of such an approach is that there is loss of sensitivity for searches in more general models which might have the same collider signatures.  In addition,  the results of  searches for one specific model would be hard to translate into other scenarios, therefore, less transportable or useful when considering a large group of new physics scenarios that might give  the same collider signatures.   There is a need for a set of simple but generalized models that can cover possible signature space or topology space which are considered by experimental searches at the LHC.

Recently, there has been rising activity in considering a simplified model approach~\cite{topologydraft}.  A simplified model is defined as a minimum set of new particles with a minimal Lagrangian to explain a particular topology (defined as a specific particle production and decay chain) and/or a particular experimental signature (defined as a particular set of final states in observed events).    A simplified model usually has a small set of model parameters, for example, masses of new particles and couplings of new interactions.  These model parameters   can be translated into production cross sections and decay branching ratios, which are often used to present   experimental search results.

It is clear that simplified models are not model-independent.  They are, however, less model-dependent compared to any specific model which is motivated by or proposed to solve certain problems.    However,   simplified models are typically   limits of specific  new physics models when heavy particles decouple.    They could also capture the characteristic features of a subset of a more complete model when irrelevant particles and interactions are removed.  
The advantage of considering a simplified model approach is that given the simple set of particles and interactions, it is easy to write down all possible event topologies and signatures.    The kinematic boundaries can be made manifest given the small set of mass parameters, which enable  the identification of kinematic ranges where existing search strategies are not efficient.   The results presented in  simplified models,  usually   in terms of masses and products  of cross section times branching ratio, can also be readily translated to a general set of more specific models, which incorporate the simplified model at certain limits.

One should bear in mind that although simplified models are usually limits of more specific, well-motivated new physics models, some of the simplified models might not correspond to any existing model.  These signature-motivated simplified models might have little physics motivation.   However, they predict clean and exotic experimental signatures that can easily be searched for at experiments.   While it is important to consider such types of models in case any positive signature appears in such channels, caution must be taken when constructing such  models.   It should also be noted that in a more complete model in which other   particles might not be completely decoupled from the simplified model sector, the phenomenology of particles in the simplified model might vary due to the existence of these additional  light states.

In this paper, we consider a simplified model approach to a same-sign dilepton $\ell^\pm\ell^{\prime \pm}$  resonance, which is an extremely clean experimental signature that has almost no Standard Model (SM) background.    The relevant ${\rm SU(3)}_Q^J$ quantum numbers  of the resonance are  ${\bf 1}_2^{0,1,2}$.   For simplicity, we only consider a spin 0 scalar in the simplified model.  The $\ell^\pm\ell^{\prime\pm}$ resonance is typically referred to as doubly charged Higgs in the literature, which is denoted as $\hpp$ \footnote{Note that even though it is usually called ``Higgs" in the literature, the neutral component of the Higgs representation that contains the doubly charged Higgs does not necessary obtain a vacuum expectation value.  It should also be noted that not all doubly charged Higgses have a neutral Higgs partner in a given representation.} .   Such a doubly charged Higgs appears in many well-motivated models, for example, in the left-right symmetric models \cite{LR}, Higgs triplet models \cite{Higgstriplet, Han:2005nk, Lee:2005kd}, little Higgs models \cite{littleHiggs}, etc., where the doubly charged Higgs typically resides in a ${\rm SU(2)}_L$ triplet.  In our simplified model approach, we consider the
three simplest cases where $\hpp$ reside in a singlet, doublet or 
triplet ${\rm SU(2)}_L$ representation.   An earlier review on   studies of doubly charged 
Higgs and the related Tevatron phenomenology can be found in Ref.~\cite{Gunion:1996pq}.

In our study, we write down the minimal Lagrangian for such a Higgs representation that includes both gauge interactions and couplings to leptons.  
In the case that $\hpp$ reside in an ${\rm SU(2)}_L$ representation with a neutral component which obtains a non-zero vacuum expectation value (vev) $\langle H^0 \rangle=v^\prime$, a $\hpp WW$ coupling arises, which is proportional to $v^\prime$.  Such a doubly charged Higgs could also be considered as a same-sign $W^\pm W^\pm$ resonance.    We discuss the phenomenological implication of such a non-zero vev when it occurs.

Same-sign dilepton resonances have been searched for at both Large Electron Positron collider (LEP), Hadron Electron Ring Accelerator (HERA), as well as the Tevatron \cite{OPAL, L3, DELPHI, OPAL_single, H1, d0limit, CDF, CDFFV}.   Null results on such searches impose a bound on 
$\sigma \times {\rm Br}$ as a function of   $\mhpp$.  A mass limit on 
$\mhpp$ for a given model can be extracted when comparing the experimental limits on $\sigma \times {\rm Br}$  with theoretical predictions.  Almost all the experimental limits on the masses are given in the framework of the Left-Right symmetric model, where $H_{L,R}^{\pm\pm}$  is part of a  ${\rm SU(2)}_{L,R}$ triplet.   We summarize all the current  experimental direct search results from various experiments, and derive limits on $\mhpp$ for the cases of singlet, doublet and triplet.   In particular, the limits on $\mhpp$ for the singlet and doublet cases have never been obtained before.  
We also summarize the direct search limits on leptonic Higgs coupling $h_{e\ell}$ as a function of $\mhpp$  based on the searches of  single production of doubly charged Higgs  at OPAL and H1 \cite{OPAL_single, H1}. 

Virtual exchange of a doubly charged Higgs  could lead to deviations from the SM prediction of Bhabha scattering, rare muon and tau decays, muon $g-2$ and muonion-antimuonion conversion.  Non-observation of these effects can be used to impose a limit on 
a combination of  leptonic Higgs couplings  and its mass: 
$h_{\ell \ell^\prime}h_{\ell^{\prime\prime} \ell^{\prime\prime\prime}}/\mhpp^2$.    Some of those constraints have been studied in the literature  \cite{Mohapatra:1992uu, Swartz:1989qz}.   These studies are, however, incomplete and outdated.  In our study, we  considered the complete set of the indirect constraints with the latest experimental bounds, and update the corresponding limits on $h_{\ell \ell^\prime}h_{\ell^{\prime\prime} \ell^{\prime\prime\prime}}/\mhpp^2$.

We also study the discovery potential of a doubly charged Higgs via same-sign dilepton resonance channel ($eeee$ and $\mu\mu\mu\mu$) at the LHC with center of mass energies,  7 and 14 TeV, respectively, for all three cases of singlet, doublet and triplet.  The LHC reach for the triplet case at 14 TeV has been studied before
at partonic level \cite{Han:2007bk}, which agrees with our results reasonably well.   We find that for the triplet case, assuming 100\% leptonic decay branching ratio of the doubly charged Higgs, a mass reach of 380 GeV (800 GeV) can be achieved for the LHC at 7 (14) TeV center of mass energy with 10 (100) ${\rm fb}^{-1}$ integrated luminosity.   The reaches for the doublet and the singlet cases are lower due to the reduced production cross sections.

The rest of the paper is organized as follows.  In Sec.~\ref{sec:model}, we describe the simplified model for the doubly charged Higgs in various ${\rm SU(2)}_L$ representations.  
In  Sec.~\ref{sec:pheno}, we discuss the productions and decays of the doubly charged Higgs at colliders.  In Sec.~\ref{sec:direct} and Sec.~\ref{sec:indirect}, we summarize the current direct and indirect search limits on the doubly charged Higgs.   In Sec.~\ref{sec:LHC}, we present the LHC reach of the doubly charged Higgs with collider analyses.   In Sec.~\ref{sec:conclusion},  we present our conclusions.

\section{Simplified Model Definition}
\label{sec:model}

The same-sign dilepton resonance can be simply modeled (in addition to the Standard Model Lagrangian) by the addition of a pair of doubly charged scalars $\hpp$. Such scalars can arise from various ${\rm SU(2)}_L$ multiplets. We consider   three simplest cases where $\hpp$ reside in a singlet, doublet or triplet ${\rm SU(2)}_L$ representation. This specification fixes the couplings of the doubly charged Higgs to SM gauge bosons.

\subsection{Representations}
We classify   $\hpp$  models by the ${\rm SU(2)}_L$ multiplet in which they appear. We consider only the cases where the $H^{++}$ is the maximally electrically charged particle in the multiplet.
%In models where there is a charged particle with charge greater than 2, for example a $H^{+++}$ this would result in a two body decay to $H^{++}$ and a $W^+$ or or alternatively a 3-body resonance in its decay to Standard Model particles.
We also make the assumption in the simplified model that there are no new particles other than the multiplet containing the $\hpp$ and no new gauge symmetries other than the SM gauge groups.

Doubly charged Higgs usually appears in  ${\rm SU(2)}_L$ triplet for almost all the models studied in the literature.  To be more general in our simplified model approach, we allow for representations under ${\rm SU}(2)_L$ other than the triplet.

\begin{itemize}
\item $H^{++}$ in a singlet $\Phi=H^{++}$: $(T=0, \ T_3= 0,\  Y = 2)$ .
\item $H^{++}$ in a doublet $\Phi=\left(
\begin{tabular}{c}
$H^{++}$\\
$H^+$
\end{tabular}
 \right)$: $(T=1/2, \ T_3= 1/2,\  Y = 3/2)$ .
\item $H^{++}$ in a triplet
$\Phi=\left(
\begin{tabular}{cc}
$H^{+}/\sqrt{2}$ & $H^{++}$\\
$H^{0}$&$-H^+/\sqrt{2}$
\end{tabular}
\right)$: $(T=1, \ T_3= 1,\  Y = 1)$ .
\end{itemize}
Here we have picked the normalization for the hypercharge $Y$ being: $Q=T_3+Y$.
%In Table.~\ref{table:rep}, we show the quantum number assignment, ${\rm SU(2)}_L$ multiplet, as well as the (three point) coupling structure of $H^{++}$ to gauge bosons and leptons.
Note that in the cases of a triplet,  there is also a neutral component $H^0$ in the multiplet.  
Once $H^0$ obtains a vacuum expectation value, it  has interesting phenomenological implications that will be discussed below.

In our discussion below for the simplified model, we assume that the doubly charged Higgs is the lightest member of the multiplet and all other components are heavy and therefore decouple from the low energy phenomenology of the doubly charged Higgs.  Note that the mass splittings between different components  of the same ${\rm SU}(2)_L$ representation are typically constrained by the electroweak precision measurements, in particular, the $T$ parameter in  the oblique parameters \cite{STU}.  Therefore,  other components can not be truly decoupled since their masses can not be
pushed to an arbitrary large value.  Most of the discussions below for  the
doubly charged Higgs  will either not be affected or can be easily modified with the existence of those states.   In our paper we will point out when extra care is needed if these states are not entirely decoupled.
  
%\begin{table}
%\begin{tabular}{|c|ccc|c|cc|c|c|} \hline
%&$T$&$T_3$&$Y$&${\rm SU(2)}_L$ multiplet $\Phi$&
%$J_\mu A^\mu$&$J_\mu Z^\mu$&$H^{++}W^-W^-$&$H^{++}\ell^-\ell^{\prime -}$ \\
%\hline
%Singlet&0&0&2& $H^{++}$& $2e$& $\frac{e}{s_W c_W}(-2 s_W^2)$&0&xxx
 %\\
%Doublet&1/2&1/2&3/2&
%$\left(
%\begin{tabular}{c}
%$H^{++}$\\
%$H^+$
%\end{tabular}
 %\right)$
%&
%$2e$&$\frac{e}{s_W c_W}{(1/2-2 s_W^2)}$&0&xxx \\
%Triplet&1&1&1&
%$\left(
%\begin{tabular}{cc}
%$H^{+}/\sqrt{2}$ & $H^0$\\
%$H^{++}$&$-H^+/\sqrt{2}$
%\end{tabular}
%\right)$
%&
%$2e$&$\frac{e}{s_W c_W}(1-2 s_W^2)$&xxx&xxx \\ \hline
%\end{tabular}
%\caption{Summary of ${\rm SU(2)}_L$ multiplets for various $H^{++}$ cases,  the neutral current interactions of $H^{++}$, as well as the
%couplings of $H^{++}$ to $W^-W^-$ and $\ell^-\ell^{\prime -}$}
%\label{table:rep}
%\end{table}

\subsection{Interactions}

 The gauge interactions of the ${\rm SU(2)}_L$ multiplet $\Phi$ that contains the doubly charged Higgs  are given by:
\begin{eqnarray}
\mathcal{L}^{\rm Gauge} &=& {\rm Tr} [ (D_\mu\Phi)^\dagger(D^\mu \Phi)]. 
\end{eqnarray}
Here $D_\mu$ is given as usual by
\begin{eqnarray}
D_\mu = \partial_\mu + i g T^a W^a_\mu + i g' Y  B_\mu.
\end{eqnarray}
The matrices $T^a$ are the hermitian generators of ${\rm SU(2)}_L$ transformations, in the representation of the $\Phi$ multiplet;  $g$ and $g^\prime$ are the usual ${\rm SU(2)}_L$ and
${\rm U(1)}_Y$ gauge couplings.  

The electromagnetic and neutral current interactions of the $\hpp$ then follow after redefining $W^3_\mu$ and $B_\mu$ in terms of $Z_\mu$ and $A_\mu$,
\begin{eqnarray}
\mathcal{L}_{\rm int}^{\rm Gauge, 3pt} = Q e  J_\mu A^\mu + \frac{e}{\sin \theta_W \cos \theta_W}(T_3-Q \sin^2 \theta_W) J_\mu Z^\mu,
\end{eqnarray}
where $Q=2$ and $J_\mu = i [H^{--}(\partial_\mu H^{++})- (\partial_\mu H^{--}) H^{++}]$.
These vertices can lead to Drell-Yan pair production of $\hpp$.
We skip the charge current interactions of $\hpp$ since it involves other components ($H^\pm$) in the multiplet that we assumed to be heavy.

The four point interactions $\Phi\Phi VV$ also arise from gauge interactions.
In particular, interactions of $H^{++}H^{--}$ with a pair of photons/$Z$s are of the form
\begin{eqnarray}
\mathcal{L}_{\rm int}^{\rm Gauge, 4pt} = \left [ (Qe)^2  A_\mu A^\mu + \left ( \frac{e}{\sin \theta_W \cos \theta_W}(T_3-Q \sin^2 \theta_W) \right)^2 Z_\mu Z^\mu \right ] H^{++}H^{--},
\end{eqnarray}
which also contribute to the pair production of $H^{++}H^{--}$ at colliders.
 
%\begin{eqnarray}
%\mathcal{L}_{int}^{Gauge, 4pt} =  \Phi^{\dagger}_k(g T^a_{kj} W_\mu^a  + g'Y \delta_{kj} B_\mu)(g T^a_{ji} W_\mu^a  + g'Y \delta_{ji} B_\mu)\Phi_i
%\end{eqnarray}
%We can drop all four point interactions involving any member of the multiplet other than $H^{++}$ if we assume that they are heavy.
Special attention should be paid to the case of the triplet, which contains a  neutral component  $H^0$.   Once $H^0$ develops  a vev  $\langle H^0 \rangle=v^\prime$, it leads to a three point coupling $g^2 v^\prime H^{++}W_\mu^-W^{\mu-}$.   Such a coupling induces single production of $\hpp$ via vector boson fusion process (VBF)
$W^\pm W^\pm \rightarrow \hpp$, as well as the decay of  $\hpp \rightarrow W^\pm W^\pm$.
$\langle H^0 \rangle$ also contributes to the masses of  $W$ and $Z$, which
leads to a tree-level deviation of the SM relation between $m_W$ and $m_Z$, or equivalently,   the electroweak $\rho$ parameter: $\rho \equiv m_W^2/(\cos^2\theta_W m_Z^2)$, which is predicted to be 1 at the tree-level in the SM.  Electroweak precision measurements thus impose a tight constraints on the value of $v^\prime$: $v^\prime \lesssim 1$ GeV~\cite{Gunion:1989in, EWPTlittlehiggs}.  
Although tree level
contributions of higher Higgs representations or vevs from multiple new
representations can be arranged to have $\rho = 1$,  one-loop contributions could, nevertheless, lead to large deviations, and fine-tuning is required to avoid large contributions.  One simple solution to avoid the $\rho$ problem is to consider representations that do not contain a neutral component, or models in which the neutral component does not develop a vev.  However, in doing so, 
$H^{++}W_\mu^-W^{\mu-}$ coupling is absent and $\hpp \rightarrow WW $ decay is not allowed.

%The WBF production of $H^{++}$ is therefore heavily suppressed
%comparing to the dominant Drell-Yan pair production. 
% some further discussion on WBF for LR model, which could be large since vR large.
%$H^{++} \rightarrow W^+W^+$, however, could still compete with $H^{++} \rightarrow \ell^+\ell^+$, even for a small value of $\langle H^0 \rangle$ \cite{Han:2007bk}.

If other members of the multiplet are lighter than the $\hpp$, new decay modes are opened.  For example if $m_{H^\pm} < \mhpp - m_W$, then the decay $\hpp \rightarrow H^\pm W^\pm$ is allowed.  
In principle, one could also allow couplings between components of the $\Phi$ multiplet arising from self coupling in the scalar potential.
This would give rise to new decays of the form $\hpp \rightarrow H^\pm H^\pm$ when kinematically allowed. The rate would depend on the vev of the neutral component $\langle H^0 \rangle$.  In our simplified model approach, we assume states other than $\hpp$ are heavy and we do not consider such decays.  Even with the existence of other light states in certain models, most of the results obtained in the simplified model still apply,  or can be easily adjusted with simple modification. 

The coupling of $\hpp$ to leptons  depends on the multiplet in which   the doubly charged Higgs appears. 
The singlet only couples to right-handed leptons; the triplet only couples
  left-handed leptons, and the doublet couples to a pair of left-handed and
right-handed ones.  The coupling structure of $\hpp$ is given below:
%Table.~\ref{table:rep}.
%\begin{itemize}
%\item Singlet: $ \mathcal{L}_{\textrm{singlet}} \supset h_{\ell\ell^\prime} H^{++} \overline{\ell_R^c}  \ell_R^\prime + h.c.$
%\item Doublet: $ \mathcal{L}_{\textrm{doublet}} \supset \frac{h_{\ell\ell^\prime}}{\Lambda}  H^{++} \overline{\ell_R^c}  \gamma^\mu \partial_\mu \ell_L^\prime + h.c. $
%\item Triplet: $  \mathcal{L}_{\textrm{triplet}}\supset h_{\ell\ell^\prime}  H^{++}\overline{\ell_L^c}  \ell_L^\prime + h.c.$
%\end{itemize}
%\vspace{1.5in}
\begin{itemize}
\item Singlet:   $h_{\ell\ell^\prime} \Phi \overline{\ell_R^c}  \ell_R^\prime + h.c.$
\item Doublet:  $\frac{h_{\ell\ell^\prime}}{\Lambda}  \overline{\ell_R^c}  \Phi^T \epsilon \gamma^\mu \partial_\mu L_L^\prime + h.c. $
\item Triplet:  $h_{\ell\ell^\prime}   \overline{L_L^c}  \epsilon \Phi  L_L^\prime + h.c.$
\end{itemize}

Here $\epsilon=i \sigma_2$ is the $2 \times 2 $ antisymmetric tensor contracting ${\rm SU}(2)_L$ gauge indices.   
$L_L^{(\prime)}=(\nu_{\ell^{(\prime)}}, \ell_L^{(\prime)})^T$ and $\ell^{(\prime)}_R$ are the ${\rm SU}(2)_L$ doublet and singlet, respectively and the $c$ in the superscript denotes charge conjugation. 
We pick the convention for the leptonic coupling $h_{\ell \ell^\prime}$ such that for off-diagonal couplings, the lower generation leptons always appear first in the associated operator\footnote{In the literature for the leptonic triplet coupling, it is sometimes defined using matrix  $h_{\ell_i \ell_j}   \overline{L_{iL}^c}  \epsilon \Phi  L_{jL} + h.c.$, where $i,j=1 \ldots 3$ for three generations.  The off-diagonal couplings $h_{\ell_i \ell_j}$ 
defined this way is 1/2 of $h_{\ell\ell^\prime}$ defined in this paper. 
}.

Note that the conjugate of a left-handed spinor is a right-handed spinor and vice-versa. Thus, in the case of the doublet, one needs an extra gamma matrix as well as $\partial_\mu$ for Lorentz invariance. The operator in this case is necessarily higher dimensional. This non-renormalizable operator could arise from integrating out  a heavy state other than the Higgs multiplets, with $\Lambda$ being the typical mass scale associated with   heavy particles.  An ultra-violet (UV) completion of the theory is needed to understand the origin of this non-renormalization operator, which is, however, beyond the scope of current study.

We included flavor mixing leptonic couplings in the interactions.  However,   $h_{\ell \ell^\prime}$ are constrained by direct and indirect searches, which will be discussed below. The couplings of $\hpp$ to a pair of quarks   are forbidden by ${\rm U(1)}_{EM}$.

\subsection{Theories with $\hpp$}
%\vspace{0.5in}
The mostly studied doubly charged Higgs is the one coming from a Higgs triplet, for example, in the left-right symmetric models \cite{LR}, Higgs triplet models \cite{Higgstriplet, Han:2005nk, Lee:2005kd}, and little Higgs models \cite{littleHiggs}.

%In the LR model, one could consider a class of models without a Higgs triplet, but the (right-handed) Higgs triplet allows for large Majorana mass terms for the right handed neutrino, $N$. The neutral component of the right-handed Higgs triplet must acquire a large vev to make the right-handed gauge bosons $W_R$s heavy. The operator $H_R N N$ (where $N$ is a doublet under $SU(2)_R$) gives rise to the large Majorana masses for the right-handed neutrinos. This naturally allows for a see-saw mechanism to explain the smallness of the observed neutrino masses. Hence, models containing a triplet are theoretically favored. See \cite{Gunion:1989in} for a more detailed discussion of this point. 
%{\bf we only interested in SU(2)L triplet here.  so whether we favor a SU(2)R triplet or not does not really matter.  do we need this paragraph of discussion?}

In the left-right symmetric model \cite{LR}, the gauge group is ${\rm SU(2)}_L \times {\rm SU(2)}_R \times {\rm U(1)}_{(B-L)}$.  In this model there are three scalar multiplets: 
 $\Phi_{LR}$ which is a bi-doublet under ${\rm SU(2)}_L \times {\rm SU(2)}_R$, as well as $H_L$ and $H_R$,  which are triplets under ${\rm SU(2)}_L$ and ${\rm SU(2)}_R$ respectively.  In particular,  the $H_L$ triplet is to be identified with the triplet $\Phi$ of the simplified model.  The introduction of Higgs triplet instead of doublet has the advantage of naturally explain the smallness of the neutrino mass through a see-saw type mechanism.  Two neutral components in $\Phi_{LR}$  obtain vevs $\kappa_1$ and $\kappa_2$, which are the dominant sources of the masses of the SM gauge bosons $W_L$ and $Z$.  The  $W_R^\pm$, which  dominantly obtain their masses via the vev of the neutral component of $H_R$, $v_R$,   are constrained to be heavy.  Therefore,  $v_R \gg\kappa_1,\ \kappa_2$. It has been shown in \cite{Gunion:1989in}  that consistency with experiments requires  $\kappa_2 \simeq 0$,  which corresponds to zero mixing between the $W_L$ and $W_R$ gauge bosons.  In addition,  the neutral component of $H_L$ could obtain a vev as well, which is denoted as $v_L$ (or $v^\prime$ in our notation).   A non-zero $v_L$ leads to deviation of the SM mass relation of $m_W$ and $m_Z$ at tree level, as indicated by the $\rho$ parameter:
\begin{equation}
\rho \equiv \frac{m^2_{W_L}}{\cos^2{ \theta_W} m^2_{Z}} = \frac{\kappa_1^2 + 2 v_L^2 }{\kappa_1^2 + 4 v_L^2}.
\end{equation}
Precision measurements constrain  the deviation of $\rho$ parameter  from 1: $|1-\rho| \lesssim 0.003 $ at the 2$\sigma$ level \cite{PDG}, which leads to $v_L \lesssim 0.04 \ \kappa_1$.
 This scenario  opens up a new production channel of the $\hpp_R$ from $W_R$ fusion since the right handed Higgs vev is not suppressed like the left handed one.   For a thorough discussion of the constraints on this model see \cite{Gunion:1989in}.
 
In the Higgs triplet model \cite{Higgstriplet, Han:2005nk, Lee:2005kd}, the interaction of the Higgs triplet with the lepton doublet is given by the $h_{\ell_i\ell_j}\bar{L_i^c} \epsilon \Phi L_j$ operator.   Once the neutral component of the Higgs triplet obtains a vev $v^\prime$, it provides a neutrino mass of 
$h v^\prime$.  To achieve a neutrino mass of around 0.1 eV, there are two possibilities: either $v^\prime$ is relatively large ($\sim1$ GeV)  with a small leptonic coupling $h\sim 10^{-10}$, or alternatively, $h \sim {\cal O}(1)$ with an extremely small $v^\prime$.  The former case requires fine tuning of the model to obtain a leptonic coupling matrix with all entries around $10^{-10}$, while the latter possibility is viable in the littlest Higgs model,  in which  a small Higgs triplet vev could be naturally obtained through the Coleman-Weinberg potential.

The Higgs triplet also appears in the Littlest Higgs model, with a global symmetry breaking pattern ${\rm SU}(5)/{\rm SO}(5)$ with an extra gauge symmetry of ${\rm SU}(2)\times {\rm U}(1)$  in additional to the SM ones. After spontaneous symmetry breaking, 14 goldstone bosons are left over, four of which are eaten by the heavy gauge bosons.   The remaining 10 scalar degrees of freedom  contain  a SM-like Higgs doublet as well as a  a complex Higgs triplet under the SM ${\rm SU(2)}_L$.   The neutral component of the Higgs triplet could develop a vev, which is tightly constrained by the electroweak precision measurements to be less than about 1 GeV \cite{EWPTlittlehiggs}.

%Higgs triplet could also appear in other Little Higgs models, for example, $SU(3)$ model with n-sites \cite{ArkaniHamed:2001nc}. Once adjoint ``chain" scalars (scalars that are adjoint under adjacent $SU(3)$ sites) obtains a vev that breaks the first site  down to the SM $SU(2) \times U(1)$, the remaining goldstone bosons are in the representation $3_0 \oplus 2_{1/2} \oplus 2_{-1/2} \oplus 1_0$, including one ${\rm SU(2)}_L$ triplet.
%{\bf but hypercharge is not correct.}
%

\section{Production and Decay}
\label{sec:pheno}

\subsection{Production}

\begin{figure}[h]
\resizebox{2.in}{!}{\includegraphics*[19,640][240, 780]{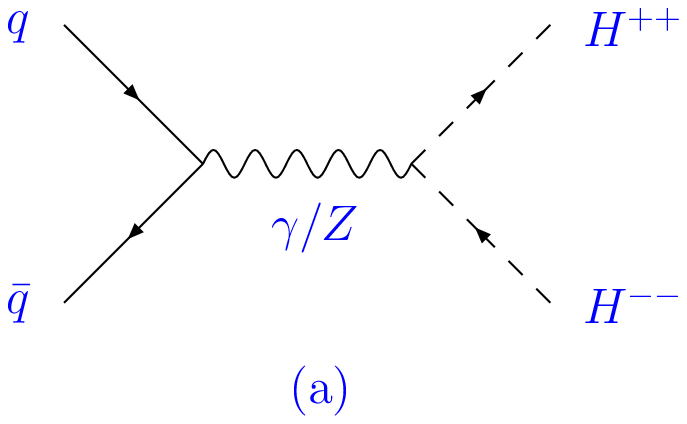}}
\resizebox{4 in}{!}{\includegraphics*[30,590][520, 730]{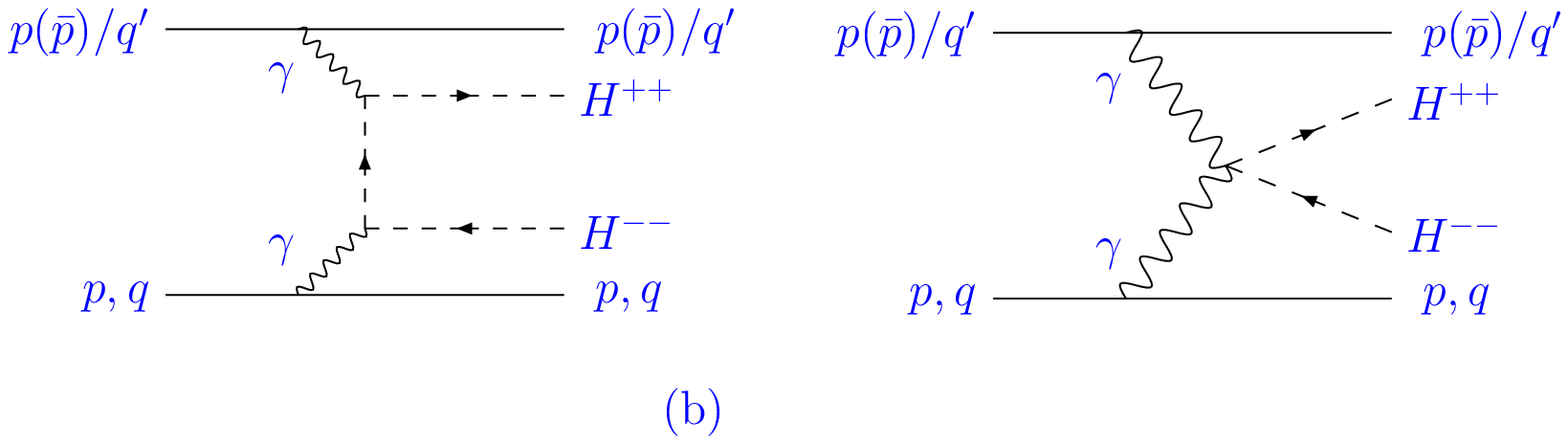}}
\resizebox{1.5in}{!}{\includegraphics* {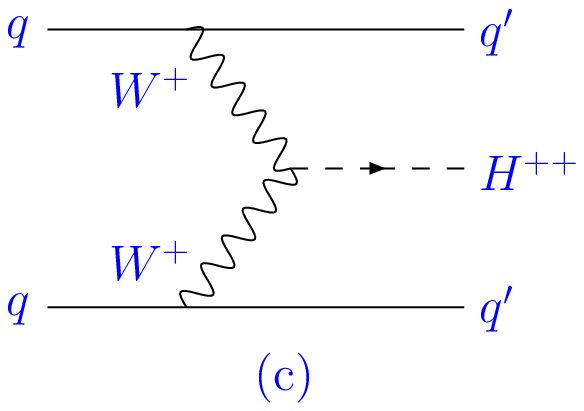}}
\hspace{0.2 in}
\resizebox{4.2in}{!}{\includegraphics*{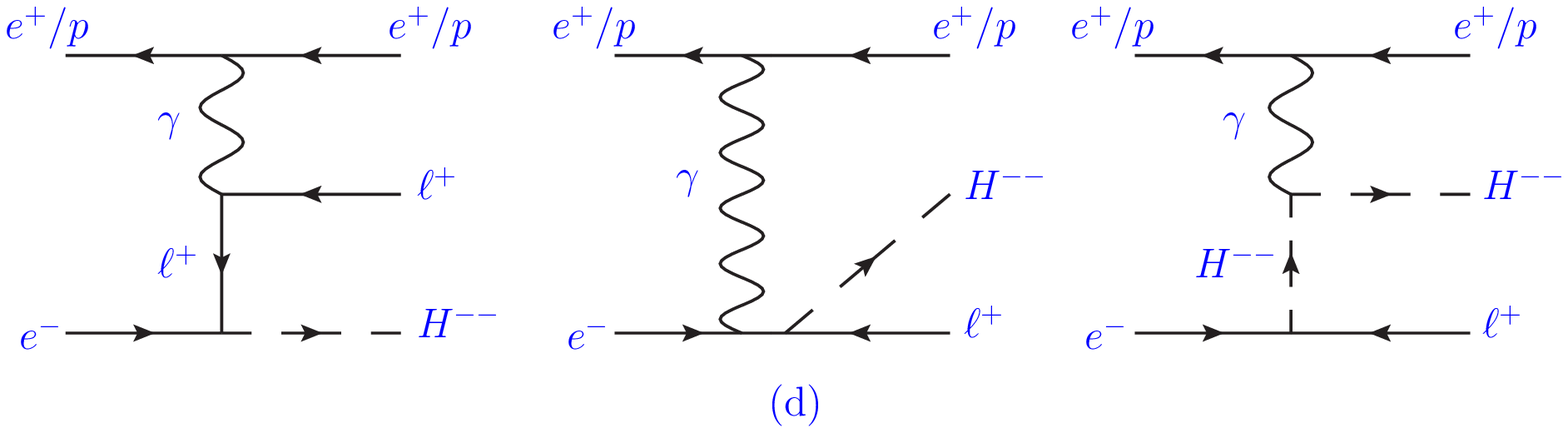}}
\caption{ Feynman diagrams for several doubly charged Higgs production channels: (a) Dominant Drell-Yan pair production process, (b) sub-dominant two photon fusion process, (c) weak boson fusion process and (d) single production via  $h_{e \ell}$ coupling. }
\label{fig:feyn}
\end{figure}

The dominant production mode for $\hpp$ is Drell-Yan pair production through an $s$-channel photon or a $Z$ boson exchange, as shown in Fig.~\ref{fig:feyn}(a). 
Since the couplings of  the doubly charged Higgs to photon and $Z$ are fixed   by the gauge structure, the pair production cross section only depends on the mass of the doubly charged Higgs for a given ${\rm SU(2)}_L$  multiplet. 
In this work we only consider the leading order (LO) cross section.  The next to leading order (NLO) QCD effects have been studied for the triplet case, giving a $K$-factor of approximately $\sim1.2 - 1.3$ for the LHC at $\sqrt{s}=14\ \tev$ with renormalization and factorization scale set to be $\mu_F^2 = \mu_R^2 = Q^2$ for the $s$-channel process. 
The variation of $K$-factor with Higgs mass is small, ranging  from 1.19 for $\mhpp=50$ GeV to 1.24 at for $\mhpp=1$ TeV,  with a peak at 1.26 for $\mhpp= 300$ GeV.
At the Tevatron, the $K$-factor varies from  $\sim1.35$ to 1.18 for $\mhpp$ between 50 GeV and 500 GeV \cite{Muhlleitner:2003me}. No NLO calculation is available for this process at the $7\ \tev$ LHC.

The two photon fusion process [shown in   Fig.~\ref{fig:feyn}(b)]  could also contribute since it has an enhancement  factor of $Q^4=16$  in the cross section compared to the case of a singly charged scalar.  
There are subdominant contributions also from $Z$ boson fusion, which is suppressed compared to the photon process.  The initial photon could be radiated from the proton (elastic process) as well from a single parton (inelastic process).  The total cross section is a combination of elastic and inelastic processes, which  is approximately 10\% of that from Drell-Yan production at the LHC (14 TeV), and always $<2\%$ of that at the Tevatron \cite{Han:2007bk}.  The addition of the two photon fusion process can be considered as a simple enhancement of the dominant Drell-Yan cross section, as our analyses does not depend sensitively on the additional hard process final states. 

With a non-zero vev ($v^\prime$) for the neutral component in the Higgs multiplet, single production of a doubly charged Higgs via   vector  boson fusion   has been studied in the literature \cite{Gunion:1989ci, WBF, Azuelos:2005uc}. 
In the simplified model, 
the contribution from VBF [see Feynman diagram in  Fig.~\ref{fig:feyn}(c)  ]  is either absent or small, since the $ \hpp WW$ coupling is proportional to $v^\prime$, which is tightly
constrained by the $\rho$  parameter, as discussed earlier in Sec.~\ref{sec:model}.   This is a general constraint on any coupling of Higgs to a pair of left-handed vector bosons and cannot be easily evaded. Some models (for example, LR models) contain a $H_R^{\pm\pm}$ [${\rm SU}(2)_L$ singlet, and part of ${\rm SU}(2)_R$ triplet], which couples to heavy gauge bosons $W_R^\pm$ with coupling coefficient proportional to $v_R$, the vev of the neutral component in the ${\rm SU}(2)_R$ triplet.  Since $v_R$ is much less constrained comparing to $v_L$, VBF production of $H_R^{\pm\pm}$ via $W_R$ might be important. 
Even in such cases,  however,  the cross sections are generically of order $\lesssim 10$ fb for relatively light scalars\,\cite{Azuelos:2005uc}. Thus we anticipate that this process is only dominant in cases of very heavy $H_R^{\pm\pm}$   when the   pair production cross section is small due to the parton distribution function suppression.   A sizeable contribution to VBF might be possible in specific models \cite{WBF}.     However, this is highly model dependent, which is less likely to be realized in general. 
 
At LEP and HERA, $\hpp$ could also be singly produced via $h_{e\ell}$ couplings, as shown in Fig.~\ref{fig:feyn}(d).  
The production cross section depends both on $\mhpp$ and $h_{e\ell}$.
Null experimental search results can therefore be used to impose constraints on $h_{e\ell}$
as a function of $\mhpp$.

In the doublet or triplet case, associated production of $H^{++}H^-$ is also possible \cite{Akeroyd:2005gt} when the $H^-$ is light as well. We do not consider such a process in the simplified model as we assume other states are heavy and decouple. 

\subsection{Decay}

$\hpp$ could decay into a pair of same-sign leptons:
$\hpp\rightarrow \ell^\pm \ell^{\prime \pm}$.  In the case of singlet and triplet with renormalizable  leptonic couplings,  the partial decay width into leptons is 
 \beq
\Gamma(\hpp \rightarrow\ell^\pm\ell^{\prime\pm})=\frac{1}{1+\delta_{\ell \ell^\prime} }\frac{|\tilde{h}_{\ell\ell^\prime}|^2\mhpp}{16\pi},\ \ \ 
 \tilde{h}_{\ell \ell^\prime}=\left\{
\begin{array}{ll}
2h_{\ell \ell^\prime}&\ \ \ \ell=\ell^\prime\\
h_{\ell \ell^\prime}&\ \ \ \ell \neq \ell^\prime
\end{array}
\right.
\label{eq:decay_Hll}
\eeq
Here,  the factor ${1}/({1+\delta_{\ell \ell^\prime} })$ counts for the phase space factor of $1/2$ for identical final state particles, and $\tilde{h}_{\ell \ell^\prime}$ accounts for the symmetry factor in the Feynman Rule.
For $h_{\ell \ell^\prime} \lesssim 10^{-7}$,  the lifetime of $\hpp$ is long enough so that it either leaves a track or appears as stable particle inside the detector.  In our discussion below, we only consider the case when $\hpp$ promptly decay  once they are produced.

Note that Eq.~(\ref{eq:decay_Hll}) does not apply to the case of a Higgs doublet, where the operator responsible for the leptonic coupling of the doubly charged Higgs is very different. 
The leptonic partial decay width, however, is highly suppressed for doublet case, since it is proportional to $(m_{\ell}/\Lambda)^2$ due to chiral suppression.    On the other hand, the competing process $\hpp\rightarrow W W$ is absent for the doublet case.    Therefore the leptonic decay branching ratio is almost 100\% 
when no other new state is present in the spectrum.

 $\hpp$ in the triplet case could also decay into $W W$ for non-zero
 $v^\prime$.  The width into $WW$ as a function of $v^\prime$  and $\mhpp$ is
\begin{eqnarray}
%\Gamma(\hpp\rightarrow W^\pm W^\pm)=\frac{g^4v^{\prime2}}{32\pi \mhpp}(3+\frac{\mhpp^4-4\mhpp^2m_W^2}{4m_W^4}\sqrt{1-4m_W^2/\mhpp^2}),
\Gamma(\hpp\rightarrow WW)&=&\frac{g^4v^{\prime2}}{32\pi \mhpp}\left(8+\frac{\mhpp^4}{m_W^4}(1-\frac{4m_W^2}{\mhpp^2})^2 \right)
\sqrt{1-\frac{4m_W^2}{\mhpp^2}} \nonumber \\
&\approx & \frac{g^4v^{\prime2}}{32\pi }
\left(\frac{8}{\mhpp}+\frac{\mhpp^3}{m_W^4}\right),
\label{eq:decay_HWW}
\end{eqnarray}
where the second line shows the dependence on $\mhpp$ in the limit of
$\mhpp \gg m_W$.  
The first term in Eq.~(\ref{eq:decay_HWW}) comes from the decay of $\hpp$ into the transverse components of $W$, which is proportional to  $1/\mhpp$. 
The second term comes from  the decay of $\hpp$ into the longitudinal  components,
which is proportional to $\mhpp^3$.  The enhancement factor of 
$({\mhpp^2}/{m_W^2})^2$ of the longitudinal modes compared to the transverse ones is governed by the Goldstone-boson equivalence theorem.

\begin{figure}[t]
\epsfig{figure=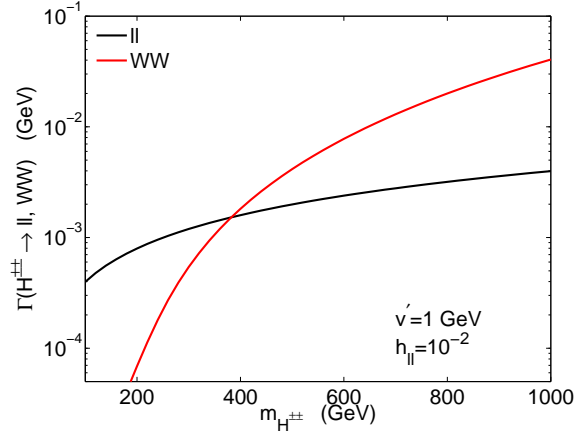, width=3in}
\caption{The partial widths for $\hpp \rightarrow WW$   and  $\hpp \rightarrow \ell\ell$  as a functions of doubly-charged Higgs mass for $v^\prime=1$ GeV, $h_{\ell\ell}=0.01$.  $\Gamma_{\ell\ell}$ and $\Gamma_{WW}$ scale as $h_{\ell\ell}^2$ and $v^{\prime 2}$ respectively.  Changes in  $v^\prime$ and $h_{\ell\ell^\prime}$ shift the two functions vertically but do not alter their shapes.   }
\label{fig:gammas} 
\end{figure}

In the case of the triplet Higgs representation, such a process competes with $\hpp\rightarrow \ell^\pm \ell^{\prime \pm}$, depending on the values of  $\mhpp$, $v^\prime$ and $h_{l l^\prime}$. 
For large $\mhpp$ the $WW$ decay becomes dominant as long as it is not strongly suppressed by the vev $v^\prime$. However,   $v^\prime$ is constrained to be less than about 1 GeV by precision measurements, resulting in an upper limit on the partial width into $W$ bosons of approximately 40 MeV at $\mhpp\sim 1$ TeV. 
In Fig.~\ref{fig:gammas}, we plotted partial decay  widths of 
$\hpp\rightarrow WW$ (lighter, red curve) and 
$\hpp\rightarrow \ell^\pm \ell^{\pm}$ (darker, black curve) for 
$v^\prime=1$ GeV, $h_{\ell\ell}=0.01$.
For this set of parameter choices,  $\hpp\rightarrow WW$ becomes dominant for 
$\mhpp \gtrsim 400$ GeV.

In the Higgs triplet model where neutrinos obtain Dirac masses via Yukawa couplings of left-handed lepton doublets with Higgs triplet \cite{Higgstriplet, Han:2005nk}, $v^\prime$ and $h_{\ell_i \ell_j}$ are 
related by   neutrino masses:
\beq
m_{\nu_{ij}}=h_{\ell_i\ell_j} v^\prime 
\eeq
Assuming  $m_\nu\sim 0.1$ eV,  for the  range of $v^\prime$ consistent with precision measurements, it is possible for the decays of $\hpp$ to be either fully dominated by leptonic final states or $WW$ final states, or a mixture of the two.

For $\hpp$ in lower representations of ${\rm SU}(2)_L$,  the decay to $WW$ does not proceed at tree level as there is no neutral component to develop a vev.  Other decays are generically possible but depend on details of the model beyond those we consider here.  In the presence of a light $H^+$ in the same multiplet,  due to the unsuppressed gauge coupling, decays to $H^{\pm(*)}W^{\pm(*)}$ would become dominant very fast once phase space is not a major concern.   
$\hpp \rightarrow H^\pm H^\pm $ could also open up when it is kinematically accessible. 
The partial decay width of this mode  depends on both $v^\prime$ and  scalar self coupling. Note that the discovery reach at the LHC that we present below is given in terms of 
$\mhpp$ and ${\rm Br}(\hpp\rightarrow \ell \ell^\prime)$, which can be applied to the cases when other decay modes are open.

\section{ Direct Search Constraints}
\label{sec:direct}

%\d0 0808.1534
The latest results on the collider direct search limits of a doubly charged scalar come from
\d0 using a dataset  of 1.1 ${\rm fb^{-1}}$ at the Tevatron Run II \cite{d0limit}.  No excess is observed for $p\bar{p}\rightarrow H^{++}H^{--}X$ with
$\hpp\rightarrow \mu^\pm \mu^\pm$.   The limit on
$\sigma(p\bar{p}\rightarrow H^{++}H^{--}X)\times {\rm Br^2(\hpp\rightarrow \mu^\pm \mu^\pm)}$ of about
20 $-$ 30 fb is derived at 95\% C.L. in the scalar mass range of 90 $-$ 200 GeV.
Assuming ${\rm Br(\hpp\rightarrow \mu^\pm \mu^\pm)}=100\%$, 
$\mhpp$ is excluded up to 143, 122  and 119 GeV
for a triplet, doublet and singlet scalar, respectively.
Note that our limit on the triplet Higgs is weaker than the \d0 published result ($\mhpp>150$ GeV at 95\% C.L.).  This is because we have not taken into account the NLO  QCD corrections to the scalar pair production cross sections, with a  $K$ factor    of about 1.35~\cite{Muhlleitner:2003me}.
 
 CDF performed a search for a doubly charged Higgs:  $p\bar{p} \rightarrow H^{++}H^{--}\rightarrow \ell^+\ell^+\ell^-\ell^-$,  in $ee$, $e\mu$ and $\mu\mu$ channels using 240 ${\rm Pb}^{-1}$ data.  A limit of $\sigma \times {\rm Br}^2$ was set to be about 30 fb for the $\mu\mu$ channel (90 GeV $<m_H<$ 150 GeV), 40 fb for the $ee$ channel (100 GeV $<m_H<$ 150 GeV), and 60 $-$ 70 fb for the $e\mu$ channel (90 GeV $<m_H<$ 150 GeV) at 95\% C.L.~\cite{CDF}.
CDF also searched for the flavor violating decay of $\hpp\rightarrow e\tau$ and $\mu\tau$ using 350 ${\rm pb}^{-1}$ data.
$\sigma \times {\rm Br^2 }$
is excluded to be about 60 fb $-$ 110 fb in the mass range of 80 $-$ 135 GeV at 95\%
C.L.~\cite{CDFFV}.
 
\begin{figure}[tb]
\centering
\epsfig{figure=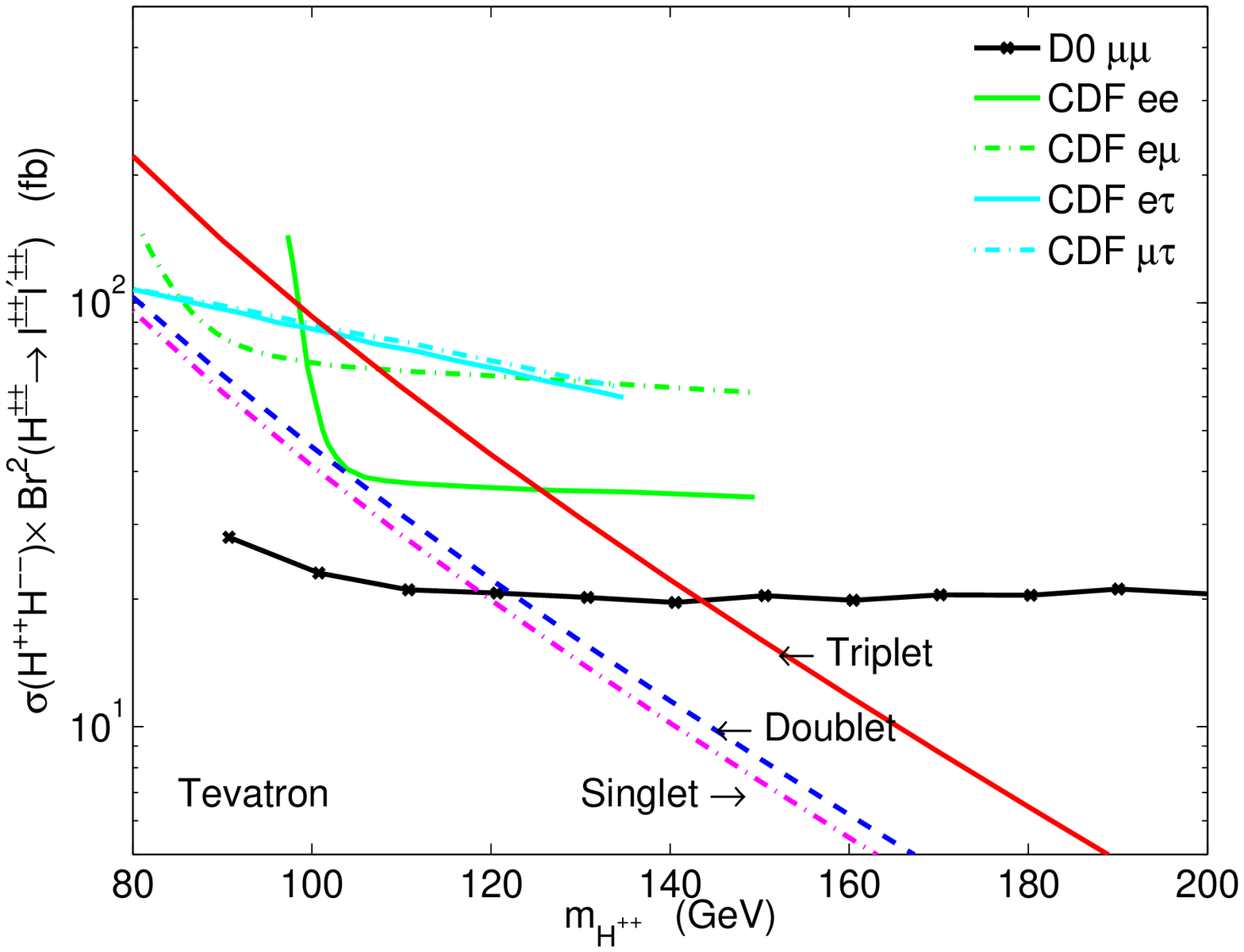, width=3in}
\epsfig{figure=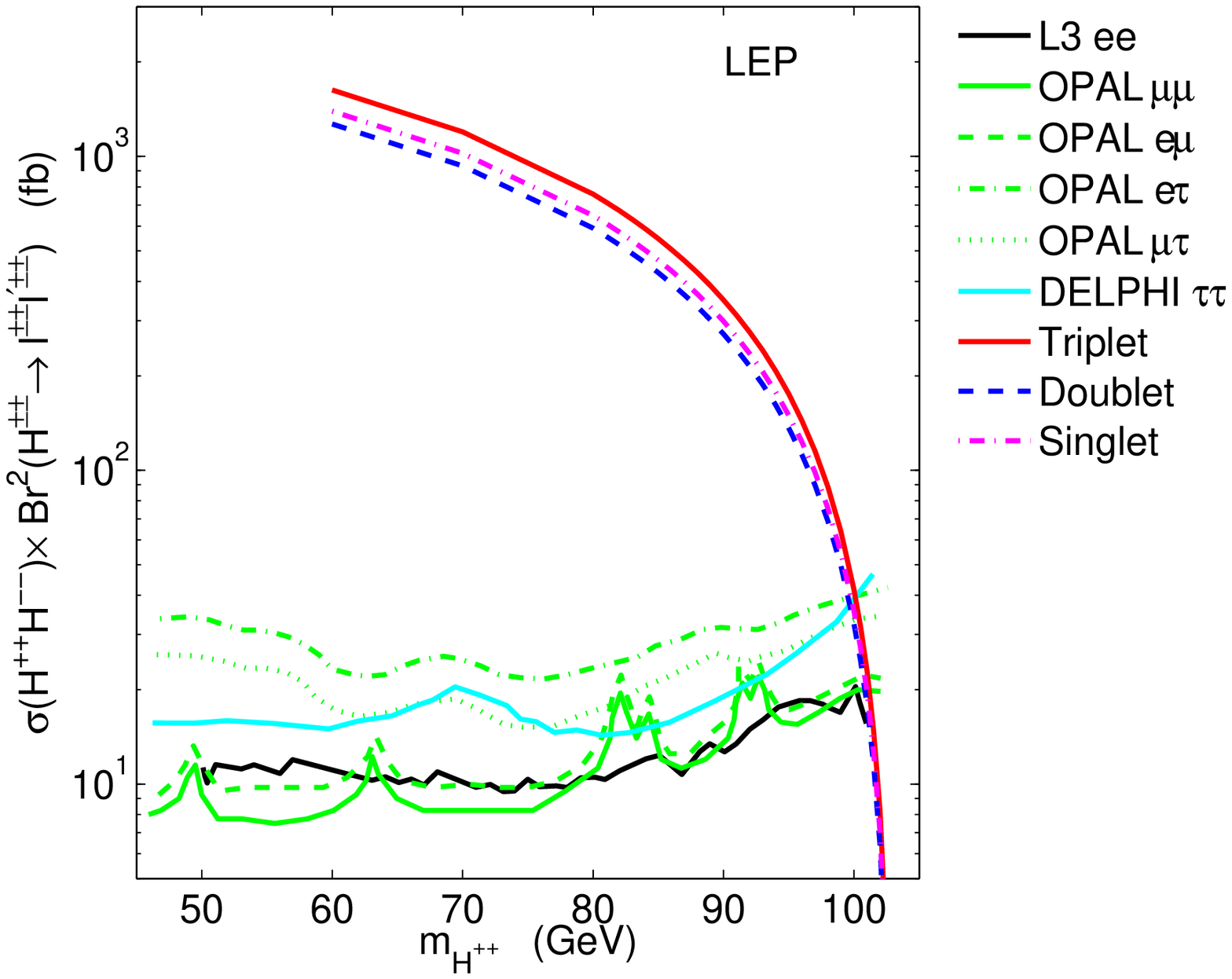, width=3in}
\caption{The current experimental direct search limits from Tevatron Run II (left panel) and LEP at $\sqrt{s}=206$ GeV (right panel).  Also shown are the Drell-Yan pair production cross sections for doubly charged Higgs in triplet, doublet and singlet multiplet.  Curves for  experimental results are taken from Refs.~\cite{d0limit, CDF, CDFFV, OPAL, L3, DELPHI} and combined. }
\label{fig:direct_pair}
\end{figure}

All the Tevatron direct search results on the pair production of doubly charged Higgses in various channels are   reproduced in the left panel of
Fig.~\ref{fig:direct_pair}, with only the best limit on a given channel being presented.
Assuming 100\% branching ratio of the Higgs decay to the relevant channel, the exclusion limits on the masses of the Higgses is  summarized in Table.~\ref{table:limit}.  Note that for doubly charged Higgs in the doublet and the singlet representation, only $\hpp\rightarrow \mu\mu$ channel  imposes   mass limits while all the other channels have not reached the required sensitivity.
LEP search results, on the other hand, could constrain $\mhpp$  in all three representations.

\begin{table}
\begin{tabular}{|c|c|cccccc|}
\hline
&&$ee$&$\mu\mu$&$\tau\tau$&$e\mu$&$e\tau$&$\mu\tau$ \\ \hline
Tevatron&Triplet&125&143 &$-$&107&103&102 \\
&(NLO)&133&150&$-$&115&114&112\\
&Doublet&$-$&122&$-$&$-$&$-$&$-$ \\
&Singlet&$-$&119&$-$&$-$&$-$&$-$ \\
\hline
LEP&T/D/S&100&100&99&100&99&99 \\
\hline
\end{tabular}
\caption{Mass limits  (in unit of GeV) of the doubly charged Higgs in the triplet, the doublet or the singlet representation from direct Tevatron and LEP searches, assuming the decay branching ratio of 100\% into the corresponding channel.    NLO results  in the table refers to the mass limits quoted in the experimental paper \cite{d0limit, CDF, CDFFV}, where NLO QCD corrections for the cross sections are taken into account. }
\label{table:limit}
\end{table}

OPAL, L3 and DELPHI     studied the pair production of doubly charged Higgses through various channels at LEP with center of mass energies
189 GeV and 209 GeV.  OPAL studied all six  $\hpp$ decay channels:  $ee$, $e\mu$, $\mu\mu$ as well as $\tau\tau$, $e\tau$, $\mu\tau$ \cite{OPAL}.  Using 614 ${\rm pb}^{-1}$ collected data, an upper limit on $\sigma\times {\rm Br}^2$ was set to be between 10 $-$ 45 fb for $\mhpp$ between 45 to 100 GeV.   The limits from similar studies at L3 with 624.1 ${\rm pb}^{-1}$ of data  is slightly worse than OPAL \cite{L3}, except the $ee$ channel.  DELPHI studied $H^{++}H^{--} \rightarrow \tau^+\tau^+\tau^-\tau^-$ channel with 570 ${\rm pb}^{-1}$ data \cite{DELPHI}.  The limits on $\sigma\times {\rm Br}^2$ are slightly better than OPAL.
All the LEP direct search results   in various channels are   reproduced in the right panel of
Fig.~\ref{fig:direct_pair}, along with the predicted pair production cross sections for the triplet, the doublet and the singlet cases at LEP with $\sqrt{s}=206$ GeV.
Assuming 100\% branching ratio of the Higgs decay to the relevant channel, the exclusion limits on the masses of the Higgses are  summarized in Table.~\ref{table:limit}.
Null results at experiments exclude $\mhpp$  to almost the LEP kinematic limit: about 100 GeV for $ee$, $e\mu$ and $\mu\mu$ channels, and 99 GeV for $\tau\tau$, $e\tau$ and $\mu\tau$ channels.

%OPAL0111059

 \begin{figure}[tb]
\centering
\epsfig{figure=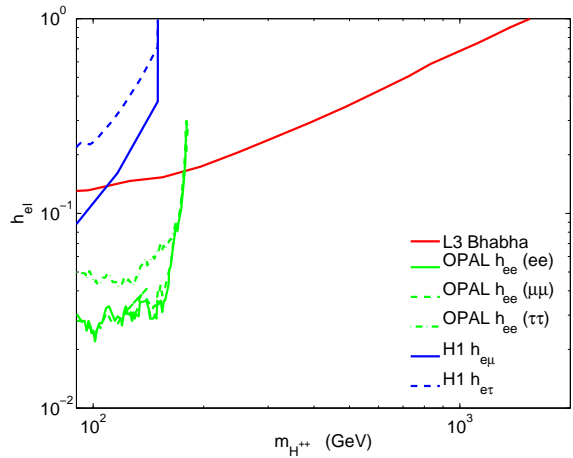, width=3in}
\caption{The current experimental limits on $h_{e\ell}$  from both direct and indirect measurements at OPAL, L3 and H1.  Curves are taken from Refs.~\cite{L3, OPAL_single, H1} and combined.}
\label{fig:hee}
\end{figure}

Single production of a doubly charged Higgs is also possible via non-zero $h_{el}$ coupling.
The corresponding Feynman diagrams are shown in Fig.~\ref{fig:feyn} (d)  for $e^+e^-$ collider and $ep$ collider.
OPAL performed a search on
$e^+e^- \rightarrow \hpp e^{\mp}e^{\mp}$ in $\hpp \rightarrow ee,\ \mu\mu, \tau\tau$ channels
with 600.7 ${\rm pb}^{-1}$ data collected at $\sqrt{s}=189 - 209$ GeV \cite{OPAL_single}.     The limits on $h_{ee}$ are shown in   Fig.~\ref{fig:hee}.  Upper limits of 0.042, 0.049 and 0.071 were set through $ee$, $\mu\mu$ and $\tau\tau$ channels for $\mhpp<160$ GeV, assuming 100\% decay branching ratio.  Note that for $\mu\mu$ and $\tau\tau$ channels, it is assumed that $h_{ee}$ coupling is small enough that the branching ratio into $\mu\mu$ or $\tau\tau$  is still 100\%.

%H1 0604027
H1 performed a search for single production of $\hpp$: $ep\rightarrow H^{\mp\mp} l^+ X$ with $H^{\pm} \rightarrow {el}$, $l=e, \mu, \tau$,  using up to 118 ${\rm pb}^{-1}$ of $ep$ data collected at HERA \cite{H1}.  Upper limits on $\sigma \times {\rm Br}^2$ are derived for $\mhpp$ between 80 to 150 GeV.  Assuming 100\% decay branching ratio, upper limits on $h_{el}$ are derived for $m_H$ up to 150 GeV.     For $h_{ee}$, the limit is weaker than the OPAL result.   For $h_{e\mu}$ and $h_{e\tau}$, the couplings are exclude up to 0.4 and 0.7 for $\mhpp=150$ GeV \cite{H1}.

\section{Indirect Searches}
\label{sec:indirect}

Contributions from virtual exchange of doubly charged Higgs in SM processes could lead to sizeable deviations from the SM predictions.  Indirect searches for $\hpp$ place important constraints on the ratio of leptonic doubly charged Higgs couplings  to the mass squared of the $\hpp$.
There are four types of processes that place indirect constraints:
\begin{enumerate}
\item Bhabha scattering,
 \item Rare decays of the muon and tau,
 \item Muonium-anti-muonium conversion,
 \item Muon $g-2$.
\end{enumerate}
Bhabha scattering is measured at high energy colliders, while the other three indirect constraints are typically low energy observables.
One important assumption we make while stating all the indirect constraints   is that no states other than $\hpp$ contribute to these processes. If other members of the multiplet  become light and contribute as well,    the constraints need to be modified correspondingly.

Below we present the indirect limits on $h_{\ell \ell^\prime}$ and $\mhpp$ (usually in terms of $h_{\ell \ell^\prime}h_{\ell^{\prime\prime} \ell^{\prime\prime\prime}}/\mhpp^2$ ) from these processes.   Note that following limits   only apply to the case of a Higgs singlet and triplet, while not applicable to the case of Higgs doublet.  For tree level processes, the contributions from $\hpp$ in the  doublet  case is suppressed by powers of small lepton masses, which can easily evade the current constraints.  For loop induced processes, the mechanism that gives rise to the non-renormalizable leptonic doublet Higgs couplings might also contribute since they could be of the same order.  The ignorance of the UV completion of the model in the simplified model approach to the doublet case makes a reliable estimation of those loop contributions very difficult, if not impossible.

\begin{figure}[h]
\centering
\scalebox{0.5}{\includegraphics*{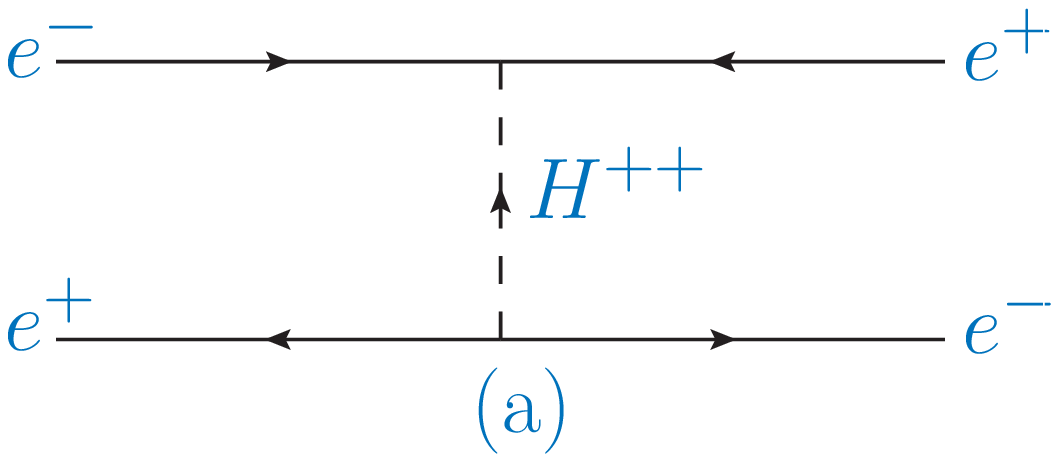}} \hspace{0.5 in}
\scalebox{0.5}{\includegraphics*{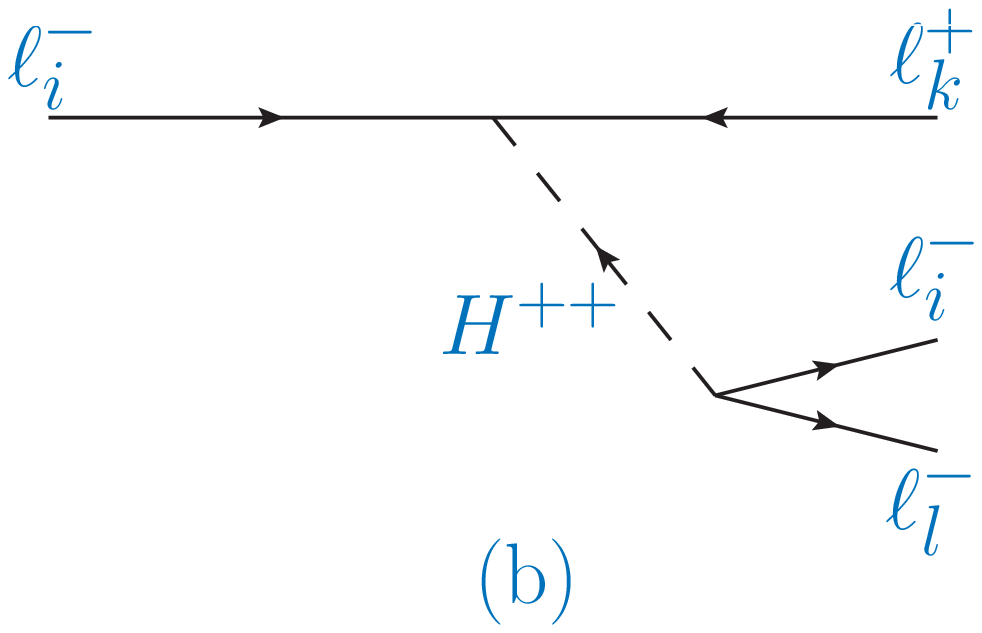}}\\
\vspace*{0.2 in}
\scalebox{0.43}{\includegraphics*{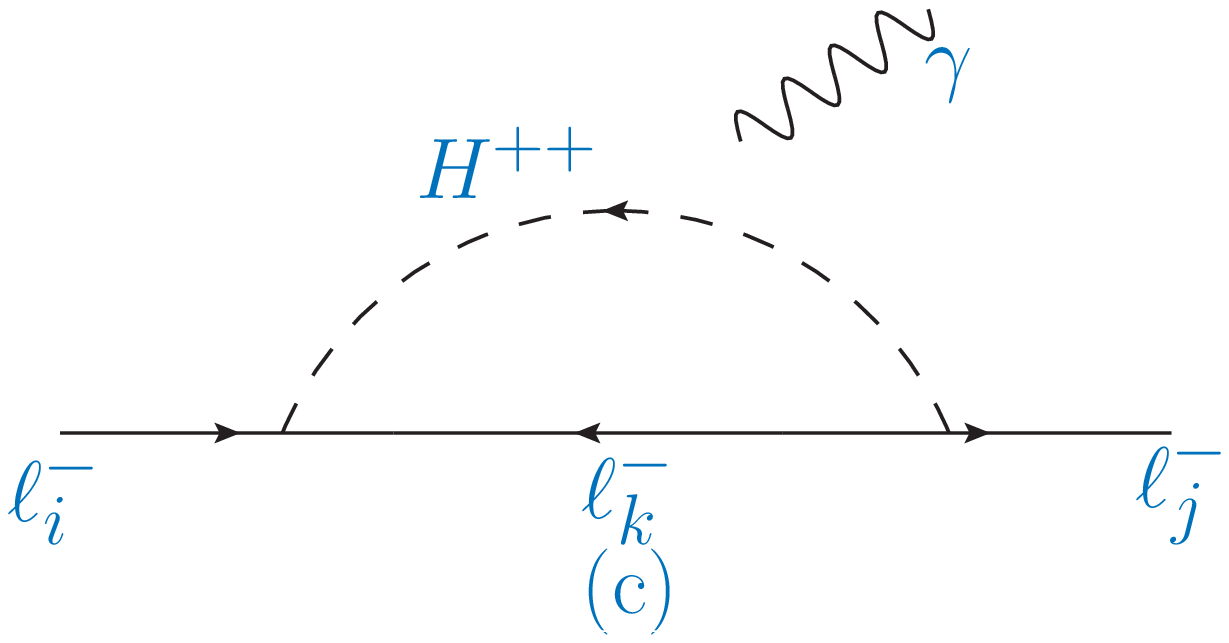}}
\hspace{0.2 in}
\scalebox{0.43}{\includegraphics*{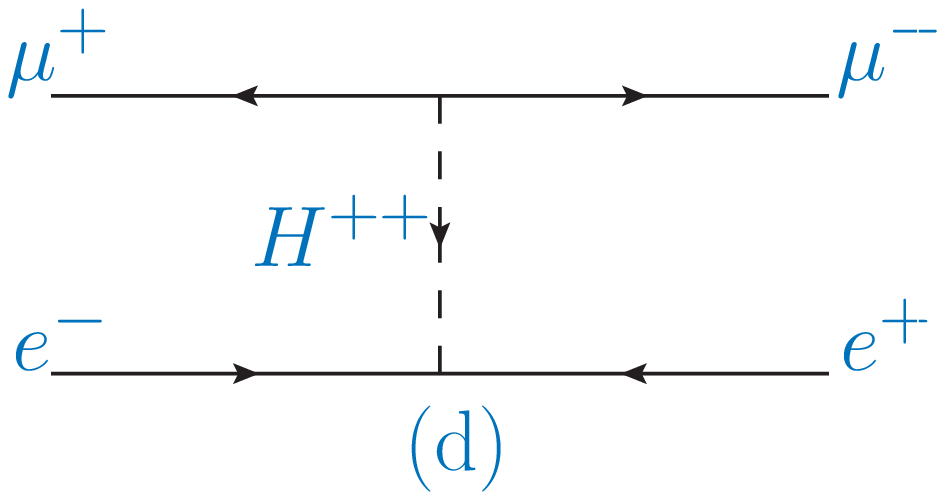}}
\hspace{0.2 in}
\scalebox{0.43}{\includegraphics*{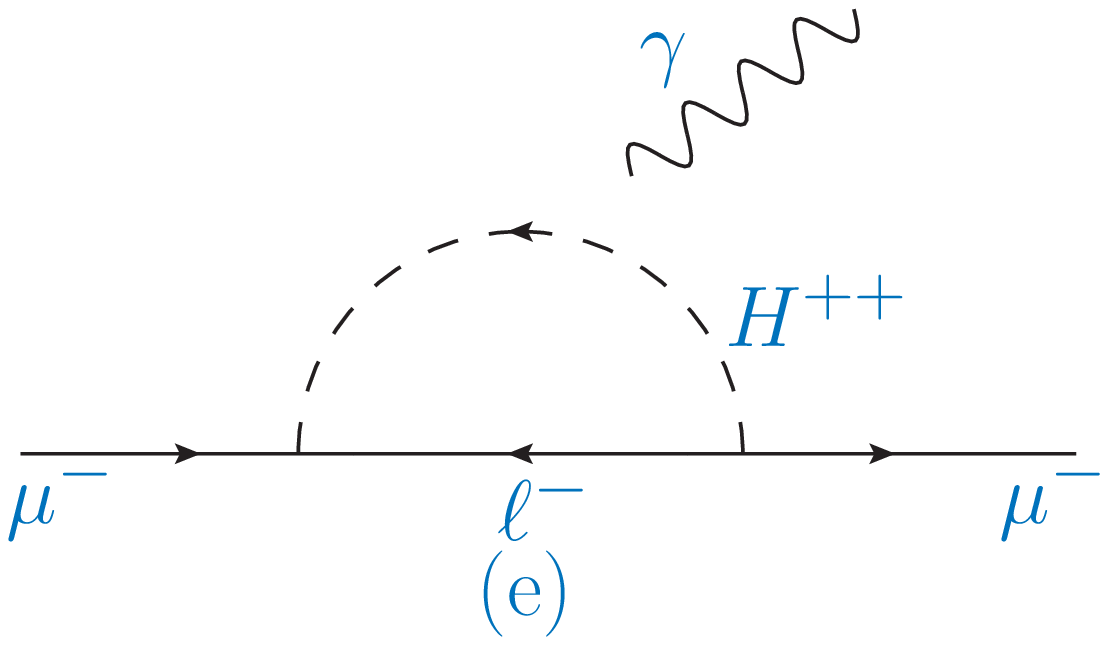}}
 \caption{Feynman diagrams for $H^{++}$ contributions to:
(a)  Bhabha scattering,
(b)  $\ell_i^- \rightarrow \ell_j^- \ell_k^+ \ell_l^-$,
(c) $\ell_i^- \rightarrow \ell_j^-  \gamma$,
(d) muonium-anti-muonium coversion,
(e) muon $g-2$. 
The detached photo line in diagrams (c) and (d) indicates that the photon can be attached to both internal charged particle lines.
  }
\label{fig:indirect}
\end{figure}

Doubly charged Higgs  contributes to Bhabha scattering $e^+e^-\rightarrow e^+e^-$  via a $t$-channel process [see Fig.~\ref{fig:indirect} (a)],  and therefore modifies the cross section and angular distribution of outgoing electrons.    OPAL derived indirect constraints on $h_{ee}$ with 688.4 ${\rm pb}^{-1}$ data collected at $\sqrt{s}=189 - 209$ GeV, ranging from 0.15 to 1.5  for $m_{H^{++}}$ between 80 GeV to 2 TeV
\cite{OPAL_single}.
L3 did a similar search with 243.7 ${\rm pb}^{-1}$ data collected at $\sqrt{s}=130 - 189$ GeV
and  446.8 ${\rm pb}^{-1}$ data collected at $\sqrt{s}=189 - 209$ GeV \cite{L3}, with results slightly better than the ones from OPAL. The indirect limit on $h_{ee}$ derived from Bhabha scattering is   also shown in Fig.~\ref{fig:hee}.  Although for the low mass region, Bhabha scattering is less sensitive compared to the direct search limits from single production of $\hpp$ via $h_{ee}$ coupling, it could probe a much heavier $\mhpp$ due to the virtual exchange of $\hpp$.

%In fact, as argued in the section where we presented the model and constraints, it was pointed out that all members of the multiplet cannot be made arbitrarily heavy without being in conflict with precision electroweak parameters. All bounds are calculated using constraint data from the 2010 PDG \cite{PDG}.

Rare decays of  $\mu$ and $\tau$ induced by  $\hpp$  were examined partly in \cite{Mohapatra:1992uu, Swartz:1989qz}.
There are two types of decays which place important constraints. The first is   tree level decay processes through   off-shell $\hpp$: $\ell_i^- \rightarrow \ell_j^- \ell_k^+ \ell_l^-$.
The second is loop induced decay:  $\ell_i^- \rightarrow \ell_j^-  \gamma$. The diagrams contributing to these decays are shown in Fig ~\ref{fig:indirect} (b) and (c).

For the tree level rare decay process $\ell^-_i \rightarrow \ell^-_j \ell^+_k \ell^-_l$,   by   comparison  with the unsuppressed decay of
 $\Gamma(\ell^-_i \rightarrow e \nu_i \bar{\nu}_e )$, we have \cite{Swartz:1989qz}  
\begin{equation}
\textrm{Br}(\ell_i \rightarrow \ell_j^- \ell_k^+ \ell_l^-)  = \frac{1}{1 + \delta_{jl}} \left |\frac{ {\tilde{h}_{\ell_k \ell_i} \tilde{h}_{\ell_j\ell_l}^\dagger}/{8 \mhpp^2} }{{G_F}/{\sqrt{2}}} \right |^2 Br(\ell_i \rightarrow e \nu_i \bar\nu_e), \ \ \ {\rm for}\ \ell_i=\mu, \tau.
\label{eq:tautree}
\end{equation}
Again, $\tilde{h}_{\ell_j\ell_l}$ = $h_{\ell_j\ell_l}$, for $j\neq l$ and $\tilde{h}_{\ell_j\ell_l}$ = $2h_{\ell_j\ell_l}$ for $j=l$.   
The factor   ${1}/{(1 + \delta_{jl})}$   
in Eq.~(\ref{eq:tautree}) accounts for the phase space factor for identical final state particles. 
 
For loop induced rare decay  $\ell_i^- \rightarrow \ell_j^-  \gamma$,  the relation between this rare decay branching fraction and the leptonic decay branching fraction is given by \cite{Mohapatra:1992uu}
\begin{equation}
\textrm{Br}(\ell_i^- \rightarrow \ell_j^-  \gamma) = \left( \frac{ \alpha}{48 \pi G_F^2} \right) \left |\frac{\tilde{h}_{\ell_i\ell_k}\tilde{h}_{\ell_j\ell_k}^\dagger}{\mhpp^2} \right |^2 \textrm{Br}(\ell_i \rightarrow e \nu_i \bar{\nu}_e ).
\end{equation}

Given the current upper limits on the rare decay branching ratios of  $\mu^- \rightarrow e^- e^+ e^-$,  $\tau \rightarrow  \ell^-_j \ell^+_k \ell^-_l$,  $\mu\rightarrow e \gamma$, $\tau\rightarrow e \gamma$, and $\tau\rightarrow \mu \gamma$, 
one can construct bounds on the ratio of couplings to $\mhpp^2$,  which are presented in Table ~\ref{table:indirect}.

Constraints from muonium-anti-muonium ($M\overline{M}$) conversion were considered in Ref.~\cite{Swartz:1989qz}.  The probability of this transition is given by
\begin{equation}
\textrm{Probability}(M \rightarrow \overline{M}) = \frac{\delta^2}{2 \Gamma_\mu^2},
\end{equation}
 where $\delta$ is   the mass difference between two states and $\Gamma_\mu$ is the total decay width of the muon.
 If one writes down an effective four fermion Hamiltonian of the form:
 \begin{equation}
 \mathcal{H}_{M\overline{M}} = \frac{G_{{M} \overline{M}}}{\sqrt{2}} {{\overline\psi }}_{{\mu }}{\gamma }^{{\alpha }}\left({1+{\gamma }^{{5}}}\right){\psi }_{{e}}{{\overline\psi }}_{{\mu }}{\gamma }_{{\alpha }}\left({1+{\gamma }^{{5}}}\right){\psi }_{{e}} + h.c., 
 \end{equation}
the mass difference can be calculated as
\begin{equation}
\delta \equiv 2\langle{\overline{M}\mid{\mathcal{H}_{{M\overline{M}}}}\mid M}\rangle=\frac {16{G}_{{M\overline{M}}}}{\sqrt{{2}}\pi {a}^{{3}}},
\end{equation}
where $a$ is the bohr radius, $a= (\alpha m_e)^{-1}$.
Using the PDG 2010 bound on $\textrm{Prob}(M \rightarrow \overline{M}) <  8.3 \times 10^{-11}$ \cite{PDG},  we get,
\begin{equation}
{G}_{{M\overline{M}}} < 0.0030 \ G_F.
\end{equation}

For $\hpp$ contribution to the $M\overline{M}$ conversion [See Fig ~\ref{fig:indirect} (d)  for the corresponding Feynman diagram], after a  Fierz re-arrangement of the amplitude,  the effective $4-$fermi coupling ${G}_{{M\overline{M}}}$ in given by 
\begin{equation}
{G}_{{M\overline{M}}} = \frac{g_{ee} g_{\mu\mu}^\dagger}{4 \sqrt{2}  \mhpp^2}.
\end{equation}
Constraints on ${G}_{{M\overline{M}}}$  allows us to place a bound on ${g_{ee} g_{\mu\mu}^\dagger}/{  \mhpp^2}$ as shown in table ~\ref{table:indirect}. 

The bounds presented in table \ref{table:indirect} show a   multi-dimensional parameter space constraint.  Each tree level process imposes a constraint on a particular product of couplings, which does not overlap.   Loop induced processes typically involve a sum of three coupling products, corresponding to three lepton flavors appearing in the loop. 
The tightest constraint comes from $\mu^- \rightarrow e^- e^+ e^- $, providing an upper limit on $h_{e \mu} h_{ee}/(\mhpp/100 {\rm GeV})^2$ to be less than $4.7 \times 10^{-7}$.   $\mu^- \rightarrow e^- \gamma $ gives the tightest constraint on 
$h_{e \mu} h_{\mu\mu}/( \mhpp/100 {\rm GeV})^2$ and $h_{e \tau} h_{\mu\tau}/(\mhpp/100 {\rm GeV})^2$, which are $2.9 \times 10^{-5}$ and $5.8 \times 10^{-5}$, correspondingly.    Tree level rare $\tau$ decays provide considerably tighter  constraints than the loop induced ones, about $(1-3 )\times 10^{-4}$ for the product of couplings to mass squared ratios.   Although much weaker, the only constraints on 
$h_{e \tau} h_{\tau\tau}/(\mhpp/100 {\rm GeV})^2$ and $h_{\mu \tau} h_{\tau\tau}/(\mhpp/100 {\rm GeV})^2$ comes from $\tau\rightarrow e \gamma$ and $\tau \rightarrow \mu \gamma$, about $3.6 \times 10^{-3}$ and $4.2 \times 10^{-3}$, respectively.    The only constraints on 
$h_{e e } h_{\mu\mu}/(\mhpp/100 {\rm GeV})^2$ comes from $M\overline{M}$ conversion, which is $1.98 \times 10^{-3}$.

\begin{table}
\begin{tabular}{|c|c|c|c|}
  \hline
  Processes & Bounds Used &Constraints \\ \hline
   $\mu^- \rightarrow e^- e^+ e^- $ & $\textrm{Br}({\mu^- \rightarrow e^- e^+ e^-}) <1.0 \times 10^{-12}$&$\frac{ h_{e e}h_{e\mu}}{\mhpp^2/(100 {\rm GeV})^2} < 4.7 \times 10^{-7} $  \\

   $\mu^- \rightarrow e^- \gamma $ & $\textrm{Br}({\mu^- \rightarrow e^- \gamma}) <1.2 \times 10^{-11}$&$\frac{ 2h_{ee}h_{e\mu} +2h_{e\mu}h_{\mu\mu}+h_{e\tau}h_{\mu\tau} }{\mhpp^2/(100 {\rm GeV})^2} < 5.8 \times 10^{-5} $  \\
\hline

  $\tau^- \rightarrow  e^- e^+ e^-$ & $\textrm{Br}({\tau^- \rightarrow  e^- e^+ e^-}) <3.6 \times 10^{-8}$ &$\frac{h_{e e}h_{e \tau}}{\mhpp^2/(100 {\rm GeV})^2} < 2.09 \times 10^{-4}$\\
 
   $\tau^- \rightarrow  e^- \mu^+e^- $ &$\textrm{Br}({\tau^- \rightarrow e^-  \mu^+e^-}) <2.0 \times 10^{-8}$ & $\frac{h_{e e}h_{\mu\tau}}{\mhpp^2/(100 {\rm GeV})^2} < 1.56 \times 10^{-4}$\\

  $\tau^- \rightarrow  e^- e^+ \mu^-$ &$\textrm{Br}({\tau^- \rightarrow  e^- e^+ \mu^-}) <2.7 \times 10^{-8}$  &$\frac{h_{e \mu }h_{e \tau}}{\mhpp^2/(100 {\rm GeV})^2} < 2.57 \times 10^{-4}$\\

  $\tau^- \rightarrow  e^- \mu^+ \mu^-$ & $\textrm{Br}({\tau^- \rightarrow e^- \mu^+ \mu^-}) <3.7 \times 10^{-8}$ &$\frac{h_{e\mu}h_{\mu\tau}}{\mhpp^2/(100 {\rm GeV})^2} < 3.00 \times 10^{-4}$\\

  $\tau^- \rightarrow  \mu^- e^+ \mu^-$ &$\textrm{Br}({\tau^- \rightarrow  \mu^- e^+ \mu^-}) <2.3 \times 10^{-8}$  &$\frac{h_{e\tau}h_{\mu\mu}}{\mhpp^2/(100 {\rm GeV})^2} < 1.67 \times 10^{-4}$\\

$\tau^- \rightarrow  \mu^- \mu^+ \mu^-$ & $\textrm{Br}({\tau^- \rightarrow \mu^- \mu^+ \mu^-}) <3.2 \times 10^{-8}$& $\frac{h_{\mu\mu}h_{\mu\tau}}{\mhpp^2/(100 {\rm GeV})^2} < 1.97 \times 10^{-4}$\\

  $\tau^- \rightarrow e^- \gamma$ &$\textrm{Br}({\tau^- \rightarrow e^- \gamma}) <3.3 \times 10^{-8}$ &$ \frac{2h_{ee}h_{e\tau} +2h_{e\tau}h_{\tau\tau}+h_{e\mu}h_{\mu\tau} }{\mhpp^2/(100 {\rm GeV})^2} < 7.2 \times 10^{-3} $  \\

$\tau^- \rightarrow \mu^- \gamma$ &$\textrm{Br}({\tau^- \rightarrow \mu^- \gamma}) <4.4 \times 10^{-8}$  &$ \frac{h_{ e \mu}h_{ e\tau} +2h_{\mu\mu}h_{\mu\tau} +2h_{\mu\tau}h_{\tau\tau}}{\mhpp^2/(100 {\rm GeV})^2} < 8.3 \times 10^{-3} $  \\
\hline 

 $M \rightarrow \overline{M}$ conversion & $\textrm{Prob}(M \rightarrow \overline{M}) <  8.3 \times 10^{-11}$ &$\frac{h_{ee}h_{\mu\mu}}{\mhpp^2/(100 {\rm GeV})^2} < 1.98 \times 10^{-3}$ \\
\hline

  Muon $g-2$ & $\Delta a_\mu^{\textrm{obs}} = a_\mu^\textrm{exp} -a_\mu^\textrm{SM} =255(63)(49) \times 10^{-11}$ & $\frac{ h_{\mu\mu}^2+\frac{1}{4}h_{e \mu }^2+\frac{1}{4}h_{\mu \tau}^2}{\mhpp^2/(100 {\rm GeV})^2} <  3.4 \times 10^{-2}$ \\
\hline
\end{tabular}
\caption{Table summarizing various processes and   indirect constraints   on the 
ratio of coupling product to mass squared.     All bounds are taken from PDG 2010 \cite{PDG} and $G_F = 1.166 \times 10 ^{-5}$ GeV$^{-2}$, $\alpha = 1/137$, and $\textrm{Br}(\tau \rightarrow e \nu_\tau \bar{\nu}_e )=17.85\%$ have been used to obtain these numerical results.   All upper limits are given at 95\% C.L. except for muon $g-2$,   
in which  constraints are derived in order to accommodate the doubly charged Higgs model at 4 $\sigma$ level, see text for details. 
For   simplicity of notation, we have ignored all the conjugations and norms of the couplings.
 }
\label{table:indirect}
\end{table}

The constraint for muon $g-2$ is slightly trickier to present in light of the observed 3.2$\sigma$ discrepancy between the SM prediction and the experimental result.
The contribution of the $\hpp$ to $a_\mu \equiv (g-2)/2 $ [shown in Fig.~\ref{fig:indirect} (e)] is  \cite{Gunion:1989in}
%\footnote{For the case of $H^+$, the contribution is $\Delta a_\mu^{H^+} = -\frac{1}{96 \pi^2}
%\frac{m^2_\mu}{\mhpp^2} \left( h^2_{\mu\mu}+h^2_{\mu e}+h^2_{\mu \tau} \right )$ }
\begin{equation}
\Delta a_\mu = -\frac{1}{6 \pi^2} \frac{m^2_\mu}{\mhpp^2} \left( |h_{\mu\mu}|^2+\frac{1}{4}|h_{e \mu }|^2+\frac{1}{4}|h_{\mu \tau}|^2 \right)=-1887 \times 10^{-11} \frac{ |h_{\mu\mu}|^2+\frac{1}{4}|h_{e \mu }|^2+\frac{1}{4}|h_{\mu \tau}|^2}{\mhpp^2/(100 {\rm GeV})^2}.
\end{equation}
 
The theoretical result for $a_\mu^\textrm{SM}$ is given by \cite{PDG}
\begin{equation}
a_\mu^\textrm{SM} =116591834(2)(41)(26) \times 10^{-11},
\end{equation}
where the errors are due to electroweak, lowest-order hadronic and higher order hadronic contributions, respectively.
%(49) if all errors are combined in quadrature.
The experimental value is given by \cite{PDG}
\begin{equation}
a_\mu^\textrm{exp} =116592089(54)(33) \times 10^{-11},
\end{equation}
where the first error is statistical and the second is systematic.
This leads to a 3.2 $\sigma$ discrepancy
\begin{equation}
\Delta a_\mu^{\textrm{observed}} = a_\mu^\textrm{exp} -a_\mu^\textrm{SM} =255(63)(49) \times 10^{-11}.
\end{equation}
Here the first error is the total experimental error and the second is the total theoretical error.

Note that the sign of this deviation from the SM is opposite to that from    $\hpp$ contribution. Thus, adding   $\hpp$ leads to an increased tension with experiment.
%We can see that the deviation from theory by $1\sigma$ constrains the couplings to satisfy
%\begin{equation}
%\frac{h^2_{\mu\mu}+\frac{1}{4}h^2_{\mu e}+\frac{1}{4}h^2_{\mu \tau}}{m_{H^{++}}^2/(100 {\rm GeV})^2} < 49/1887 = 0.0259
%\end{equation}
%Alternatively we have a tension with experiment.
%To formulate a constraint, we first define a new $\Delta a_\mu^{\textrm{observed}}$,
%\begin{equation}
%\Delta a_\mu^{\textrm{observed}} \equiv a_\mu^\textrm{exp} -a_\mu^{\textrm{SM}+H^{++}} =\left ( 255 + 1887 \frac{h^2_{\mu\mu}+\frac{1}{4}h^2_{\mu e}+\frac{1}{4}h^2_{\mu \tau}}{m_{H^{++}}^2/(100 {\rm GeV})^2} \right ) (80) \times 10^{-11}.
%\end{equation}
Requiring  theory and experiment to be consistent within 4$\sigma$ gives: 
\begin{equation}
\frac{ |h_{\mu\mu}|^2+\frac{1}{4}|h_{e \mu }|^2+\frac{1}{4}|h_{\mu \tau}|^2}{\mhpp^2/(100 {\rm GeV})^2} <  0.034.
\end{equation}
Given the existing tension between the SM prediction and the measured value of muon $g-2$, we take the attitude that other new physics contributions which give rise to positive $\Delta a_\mu$ can explain this discrepancy and allow for a $H^{++}$ with greater confidence.

\section{Collider Studies}
\label{sec:LHC}
 
In this section, we study the LHC discovery reach  for a doubly charged Higgs with center of mass energies 7 TeV and 14 TeV, respectively.  The signal is  pair production of doubly charged Higgses with the subsequent decay of $\hpp \rightarrow \ell^\pm \ell^{ \pm}$.  In our analyses, we only assume flavor diagonal decay and focus on   $e^+e^+ e^-e^-$ and $\mu^+ \mu^+ \mu^-\mu^-$ final states.   The results for flavor off-diagonal decay $e^+\mu^+ e^-\mu^-$ are very similar.   We did not consider $\tau$ leptons as their identification at hadron colliders is much more difficult.  The dominant SM background is   four lepton final states from 
$(Z/\gamma^*)(Z /\gamma^*)$.
We generate both signal and background events with MadGraph/MadEvent 4.4.56\,\cite{Alwall:2007st}, use PYTHIA 6.4.20\,\cite{Sjostrand:2006za}
to simulate showering, hadronization, and underlying event effects, and PGS 4\,\cite{pgs}
tuned to match the ATLAS experiment for detector simulation. Our event
generation includes only the dominant  Drell-Yan  pair production,
and a  NLO $K$-factor is not included in our numerical results below. 

\begin{figure}[ht]
  \begin{center}
    \subfigure{
      \includegraphics[width=.47\textwidth]{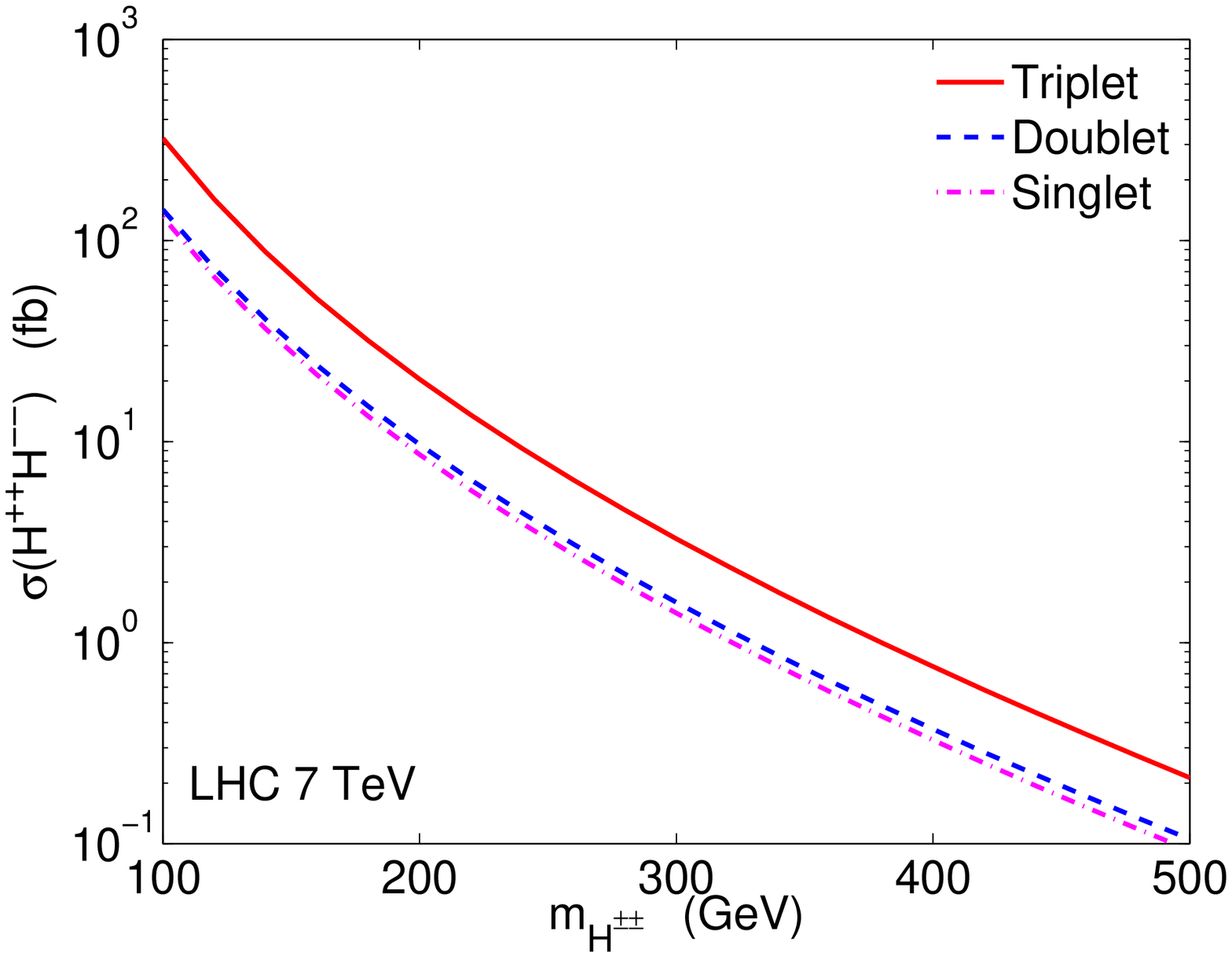}
    }
    \subfigure{
      \includegraphics[width=.47\textwidth]{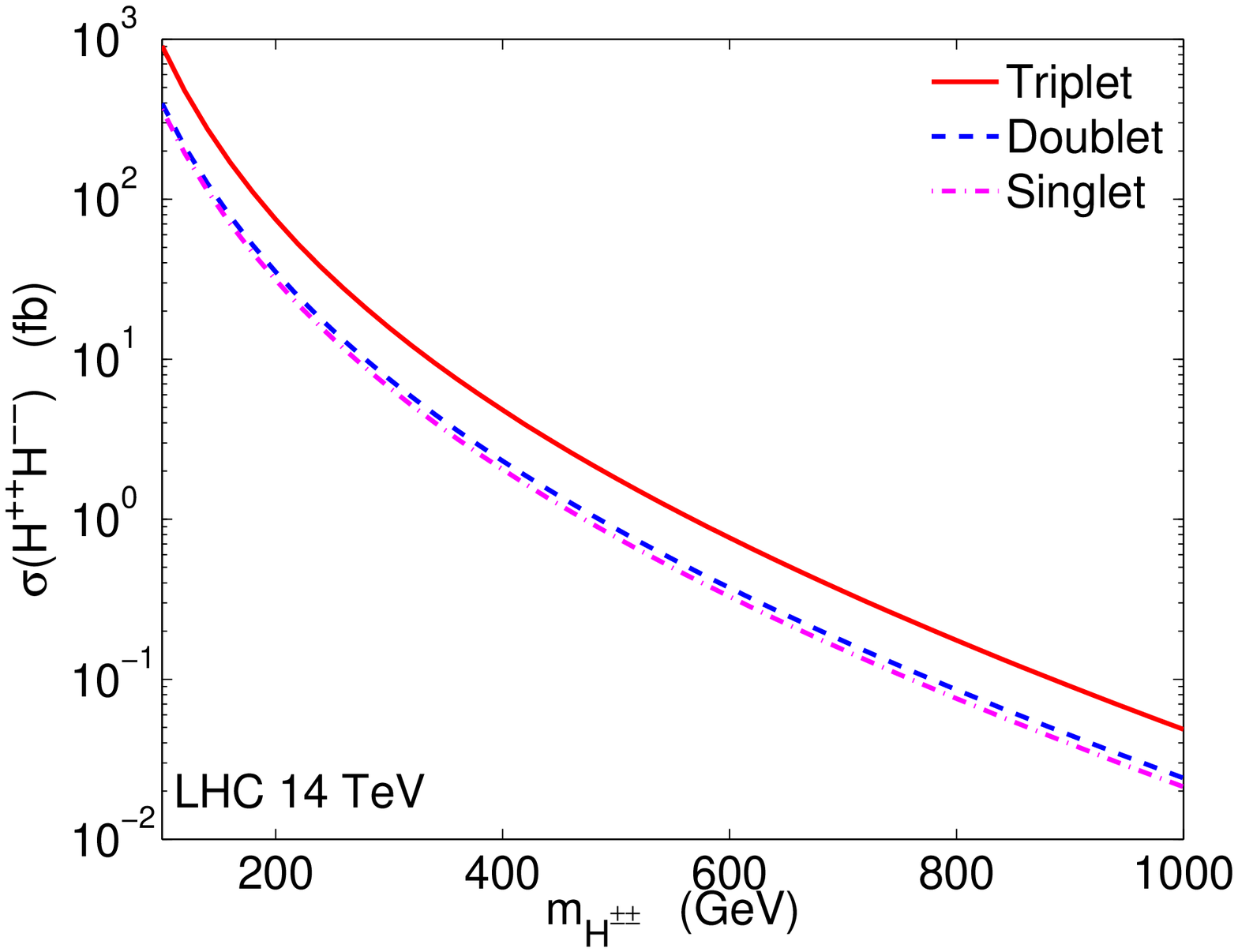}
    }
  \end{center}
  \caption{\label{fig:LHCsigmas} Leading-order Drell-Yan cross sections for $H^{++}H^{--}$ pair production at the LHC. The left panel shows cross sections for $\sqrt{s}=7\ \tev$ and the right one for $\sqrt{s}=14\ \tev$.  }
  \end{figure}

Fig.~\ref{fig:LHCsigmas} shows the tree level Drell-Yan pair production cross sections for doubly charged Higgs at the LHC with center of mass energies 7 and 14 TeV.   At the 7 TeV LHC the cross sections range between 300 fb and 0.1 fb for masses between 100 and 500 GeV.    
At the 14 TeV LHC, the cross section ranges from 1 pb to 0.02 fb for $\mhpp$  in the  range of 100 $-$ 1000 GeV.  The difference of the cross sections for the cases of ${\rm SU}(2)$ triplet, doublet, or singlet comes from the isospin dependence of the $Z$ couplings. 
The triplet case is the most copiously produced because the couplings of the $Z$ boson and photon are such that constructive interference between   two amplitudes occurs in that case. The doublet case has a much smaller amount of constructive interference between the photon and $Z$ boson, and the singlet case has a larger destructive interference effect. 
The Drell-Yan cross section for the Tevatron and LEP can be found in 
Fig.~\ref{fig:direct_pair}   when we discuss the constraints from direct detections. 
 
We select from
the reconstructed events only those which contain at least four leptons and
impose the following cuts:

\begin{enumerate}[{\bf I}]
\item At least four isolated electrons or muons with $|\eta_\ell |<2.4$.
\item Exactly four leptons with $p_{T\ell}>15$ GeV,  leading lepton with  $p_{T\ell_1}>30$ GeV.  
\item  No jets within $\Delta R_{j \ell}=0.5$ of any of the selected leptons.    
\item  $\met<25$ GeV.
\item No pair of oppositely-charged leptons with $|m_{\ell\ell}-m_Z|<15$ GeV.
 Each pair of same-sign leptons must reconstruct to the same invariant mass within 5\%.
\end{enumerate}

The requirement that only 4 leptons pass the kinematic cuts is imposed to simplify the analysis, as it becomes clear which pairs would be due to the $\hpp$ decays. Isolation of leptons from jets is imposed to avoid so-called fake leptons from heavy flavor decays or mis-identification of light flavor jets. The cut on $\met$ very efficiently rejects events where one or more leptons come from $W$ bosons,  which are usually accompanied by large missing $E_T$.  The invariant mass cuts reject events including a $Z$ boson and ensure that the events fit the expectations we have of resonant production in the same-sign dilepton channel.

The acceptance after these various cuts is shown for different values of $\mhpp$
in table\,\ref{tab:7TeV} for $\sqrt{s}=7$ TeV.  After requiring at least four leptons be reconstructed by PGS, which removes
about $40\% - 50\%$ of signal events, no significant acceptance loss is incurred before  the
invariant mass cuts. 
The low cut efficiency at Cut I  is in part due to normal identification efficiencies, isolation, and eta requirements for the leptons,  and in part due to the likely difficulty of identifying two distinct lepton tracks which are collinear and same sign.
The efficiency of the $m_{\ell\ell}$ cut is highly dependent on the mass of the $\hpp$ and varies from accepting about half to about 90\% of the 
remaining events. This mass dependence is largely due to better lepton energy and $p_T$ resolution $\Delta E/E$,  $\Delta p_T/p_T$ for more energetic leptons.  
The total efficiency ranges from about  20\% for low mass to about 50\% for
$\mhpp$ around 400 GeV.  
These cuts are, however, very efficient in rejecting SM
backgrounds for this search. We simulated the SM 4-electron background using
identical tools and found that these cuts have a background acceptance of only $0.026\%$.  Using the tree level cross section of 27 fb, the final cross section is only 0.007 fb, which is negligible compared to the signal.

\begin{table}
\begin{tabular}{|r|c|c|c|c|c|c|c|}
\hline
$\mhpp$ & $\sigma_0^{\rm triplet}$  (fb) & $\geq4$ leptons & $p_{T\ell}$ cuts & $\Delta R_{j\ell}>0.5$ & $\met<25\gev$ & $m_{\ell\ell}$ cuts & $\sigma_{\rm after}^{\rm triplet}$ (fb)\\ \hline
100 \gev & 324 & 52\% & 44\% & 43\% & 43\% & 21\% & 68 \\ \hline
200 \gev & 20.0 & 57\% & 55\% & 55\% & 55\% & 40\% & 8.0 \\ \hline
300 \gev & 3.1 & 58\% & 57\% & 57\% & 56\% & 46\% & 1.4 \\ \hline
400 \gev & 0.71 & 58\% & 57\% & 57\% & 55\% & 50\% & 0.36 \\ \hline
\end{tabular}
\caption{Cumulative efficiencies of cuts on electron signal events at $\sqrt{s}=7\ \gev$.  Cross section for the triplet case is given as an example.   
}
\label{tab:7TeV}
\end{table}

\begin{figure}[ht]
  \begin{center}
    \subfigure{
      \includegraphics[width=.47\textwidth]{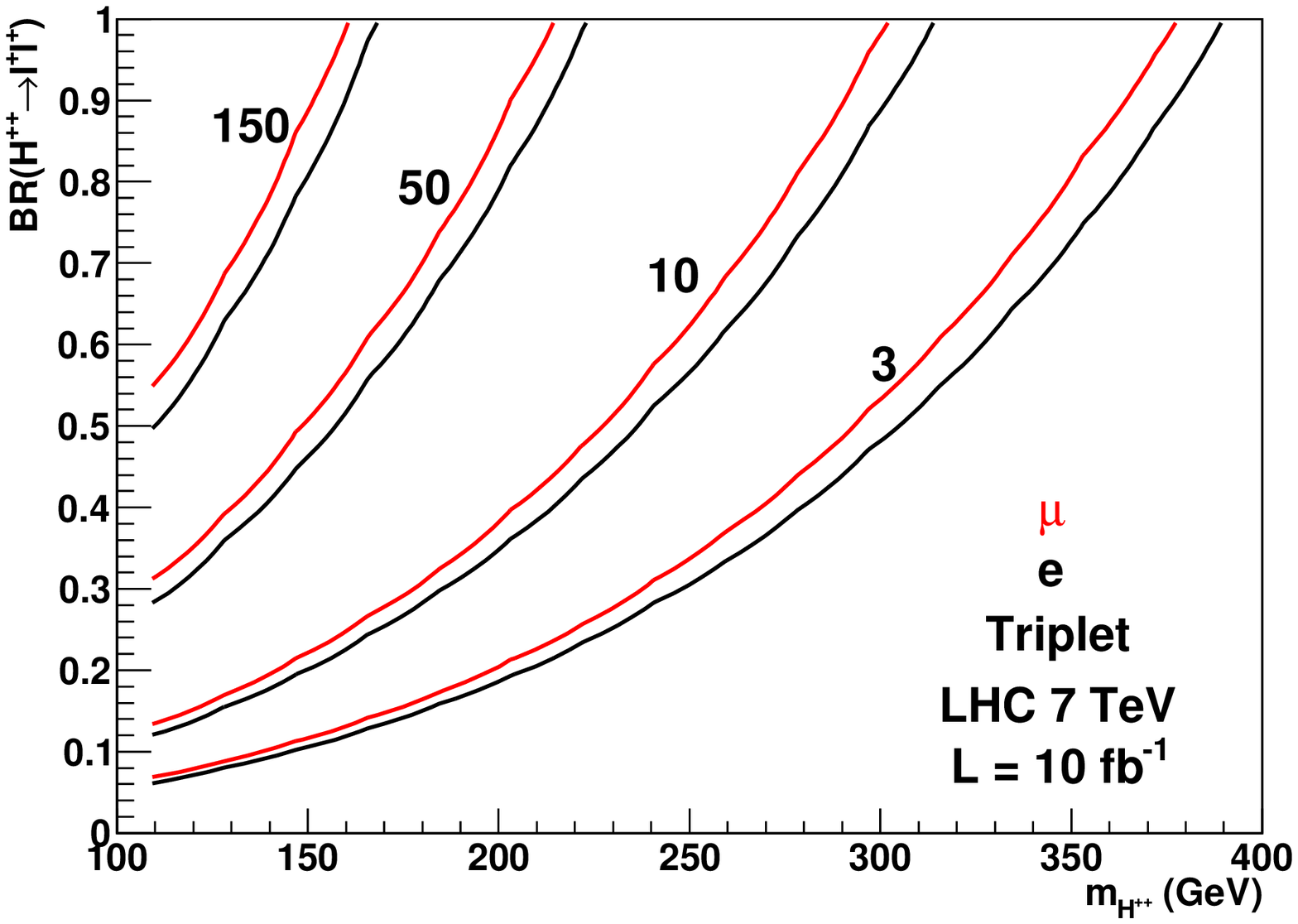}
    }
    \subfigure{
      \includegraphics[width=.47\textwidth]{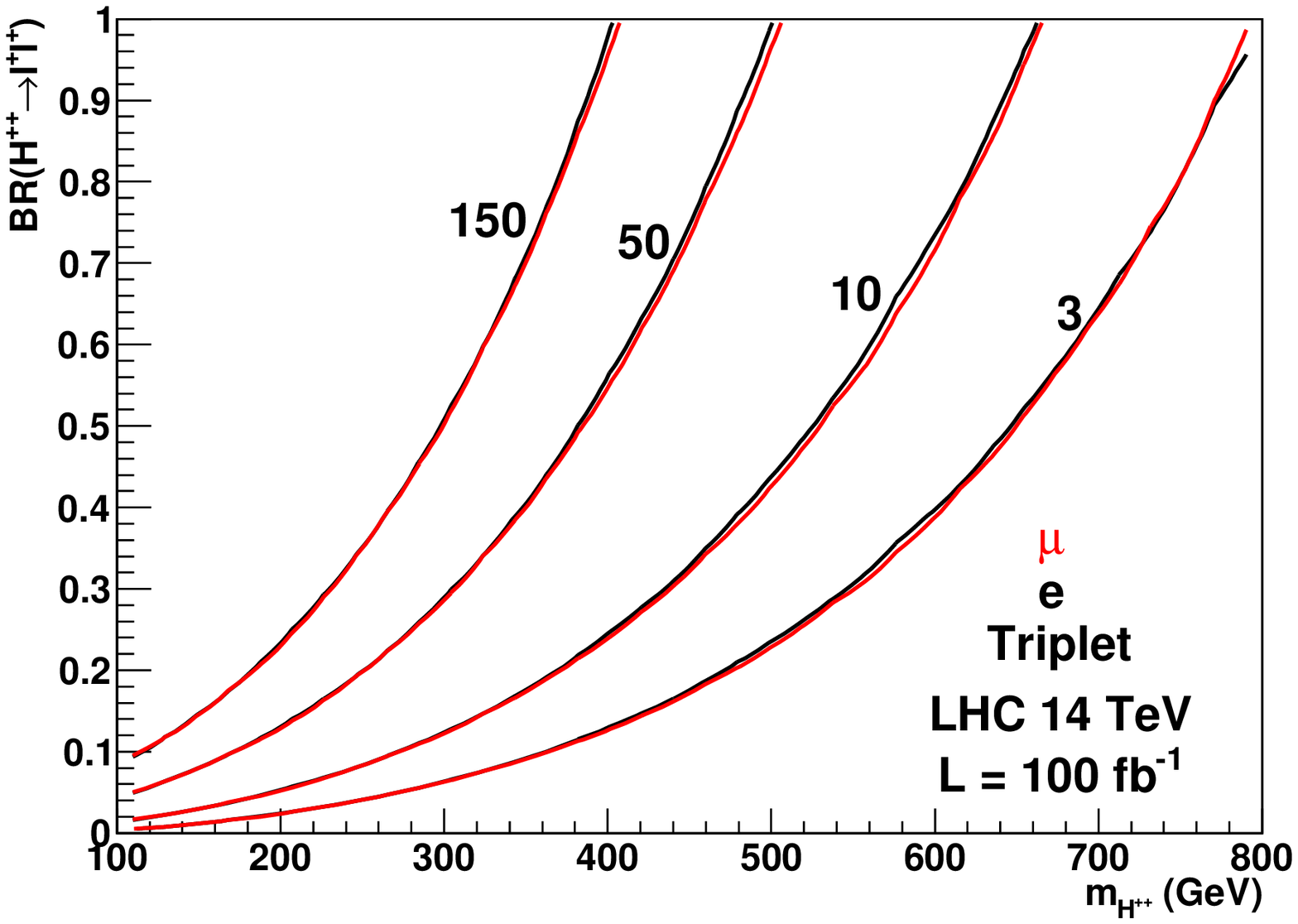}
    }
    \subfigure{
      \includegraphics[width=.47\textwidth]{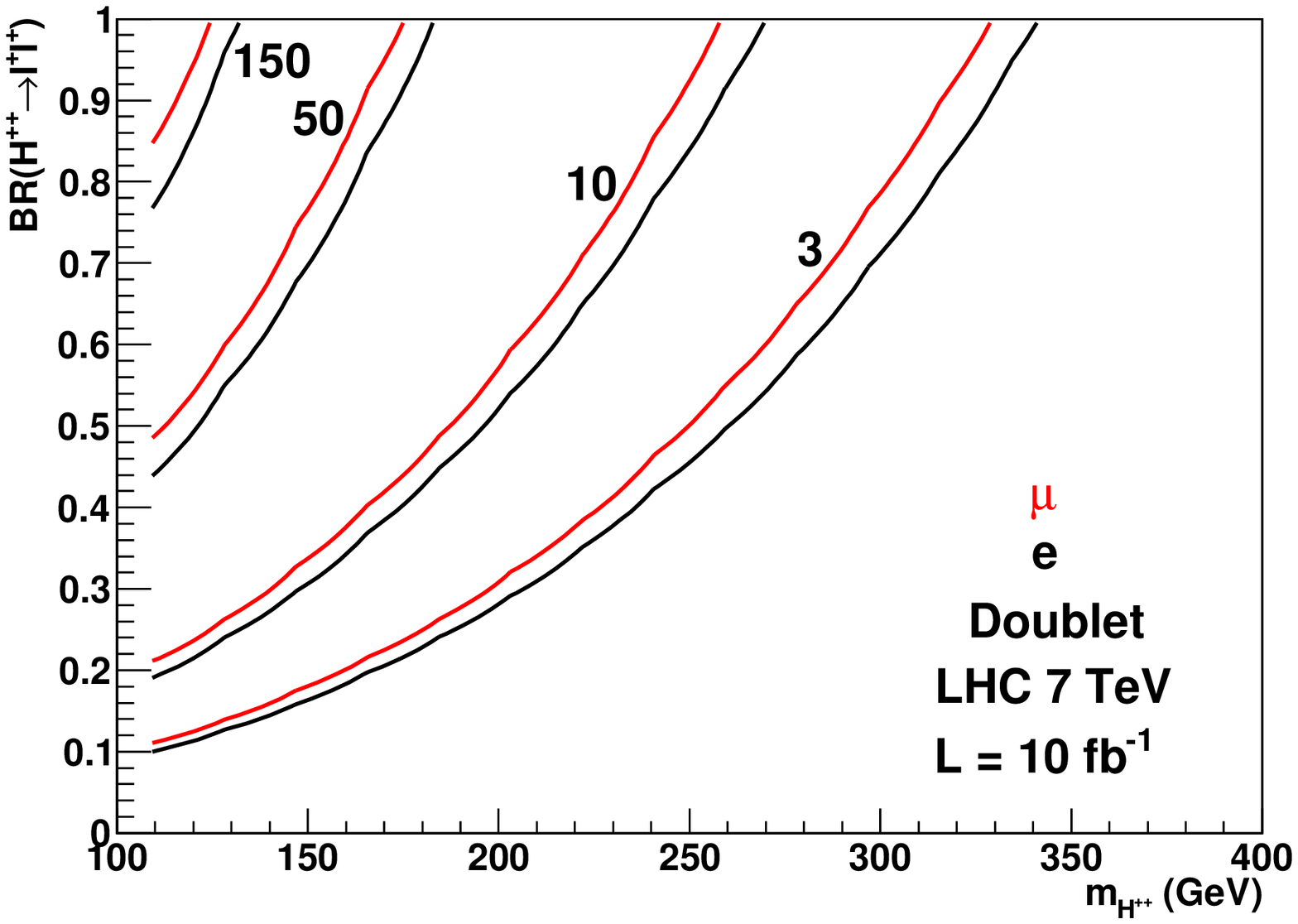}
    }
   \subfigure{
      \includegraphics[width=.47\textwidth]{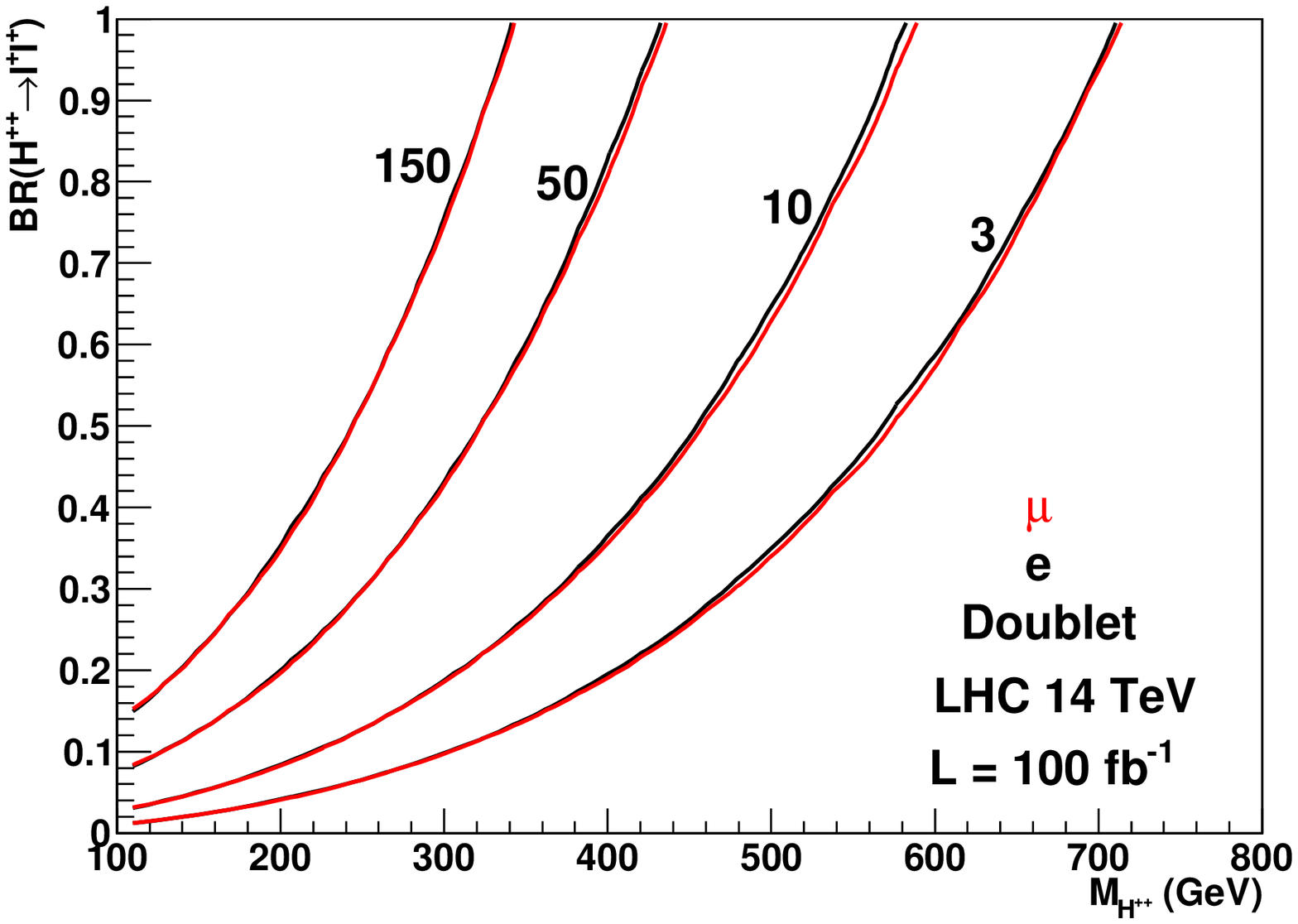}
    }
     \subfigure{
      \includegraphics[width=.47\textwidth]{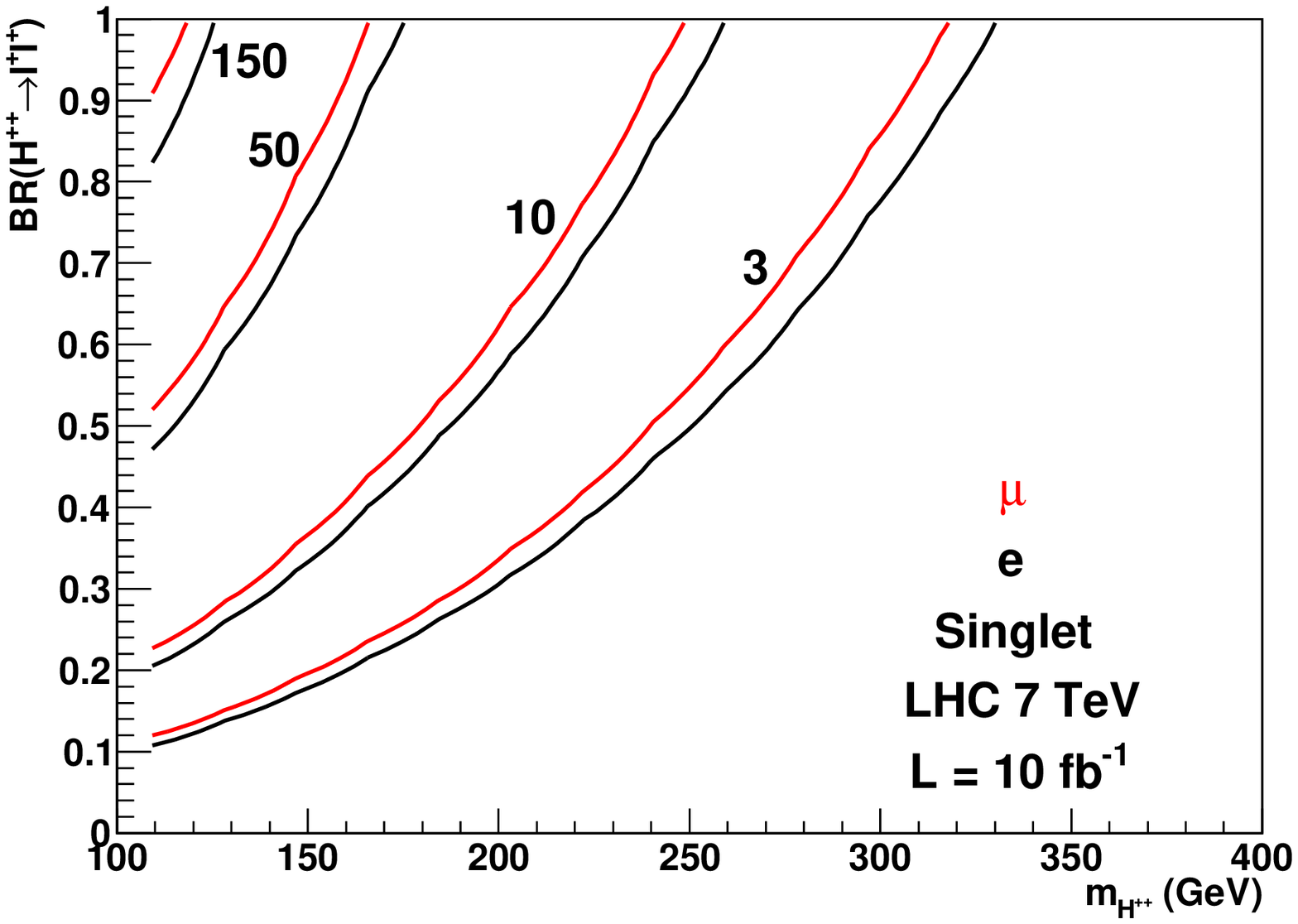}
    }
     \subfigure{
      \includegraphics[width=.47\textwidth]{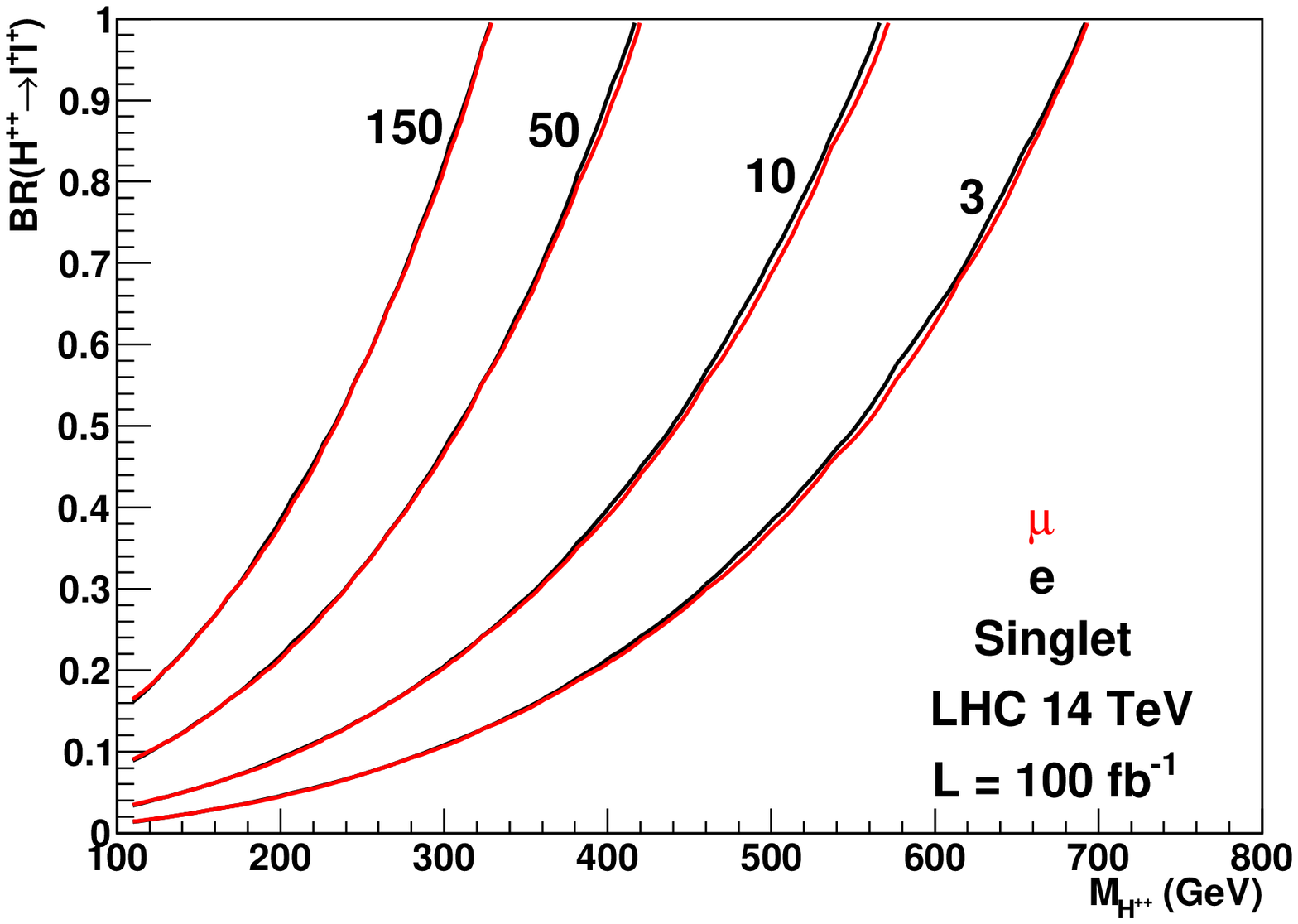}
    }
 \end{center}
%  \label{fig:7TeV}
  \caption{\label{fig:7TeV} Contours of number of events expected to pass all cuts for the Drell-Yan pair production of doubly charged Higgses with decays to $ee$ (black/dark contours) or $\mu\mu$ (red/light contours). 
 These figures assume 10 fb$^{-1}$ of integrated luminosity at LHC 7 TeV (left column) and 100 fb$^{-1}$ of integrated luminosity at LHC  14 TeV (right column).}
 \end{figure}

As the search is effectively background-free, the only limitation to search
sensitivity is luminosity. In the left column of Fig.~\ref{fig:7TeV}, we   plotted the expected number of events for
the three ${\rm SU}(2)_L$ representations  in parameter space of $\mhpp$
  and ${\rm Br}(\hpp\rightarrow \ell\ell)$, $\ell=e,\mu$  for the LHC running at 7 TeV with $10\ {\rm  fb}^{-1}$ of data.  We are capable of
excluding or seeing strong evidence (3 accepted events) of doubly charged scalar   with
masses up to 330 GeV for the least favorable singlet case and up to 380 GeV for
the triplet case, assuming a 100$\%$ branching ratio to leptons (electrons or
muons). These searches will significantly extend the current limits on such
particles.   
 The samples we analyzed
considered only
flavor-diagonal decays into same-sign electrons or muons, but the flavor
non-diagonal cases ($\hpp\rightarrow e\mu$) should differ only by a slight change in efficiency for the
reconstruction of the leptons.

\begin{table}
\begin{tabular}{|r|c|c|c|c|c|c|c|}
\hline
$\mhpp$ & $\sigma_0^{\rm triplet}$  (fb) & $\geq4$ leptons & $p_{T\ell}$ cuts & $\Delta R_{j\ell}>0.5$ & $\met<25\gev$ & $m_{\ell\ell}$ cuts & $\sigma_{\rm after}^{\rm triplet}$ (fb)\\ \hline
200 \gev & 75 & 50\% & 49\% & 47\% & 46\% & 30\% & 24\\ \hline
400 \gev & 4.7 & 57\% & 57\% & 55\% & 40\% & 35\% & 1.6 \\ \hline
600 \gev & 0.73 & 59\% & 59\% & 57\% & 29\% & 27\% & 0.20 \\ \hline
800 \gev & 0.16 & 60\% & 60\% & 58\% & 22\% & 21\% & 0.034 \\ \hline
\end{tabular}
\caption{\label{tab:14TeV}Cumulative efficiencies of cuts on electron signal events at $\sqrt{s}=14$ GeV. Cross section for the triplet case is given as an example.  
}
\end{table}

For a study at the 14 TeV LHC,  we applied identical cuts, with efficiencies of signal events as shown in table\,\ref{tab:14TeV}.    Unlike the 7 TeV case,  a significant fraction
of events simulated for high mass scalars fail the missing energy cut, with
the proportion reaching almost $60\%$ for scalars of masses around 1 TeV.
This is due largely to the additional radiation associated with higher energy events 
as well as large uncertainties in the mismeasurements of $p_T$ for more energetic objects.
Both effects contribute to the low efficiencies of signal events passing the relatively low $\met$ requirement, especially for the high $\mhpp$ region.

%While the cut on isolating leptons from jets  (Cut III, $\Delta R_{}$ cut) is largely irrelevant for signal events at 7 TeV, the same cut rejects certain amount of events at 14 TeV due to the increased likelihood of initial state radiation off of harder beams. 

Similar to the 7 TeV case, the cuts are 
sufficient to reject all SM backgrounds.   
The event contour for 100 fb$^{-1}$ in shown in the right column of Fig.~\ref{fig:7TeV}. 
We find that the 14 TeV LHC is sensitive to
boson masses below 700 GeV for the singlet case and up to 800 GeV for the triplet
case, again assuming a 100 $\%$ branching ratio to leptons.

We note that our results differ from those obtained in\,\cite{Han:2007bk} for the triplet case,
which is a parton level study with smearing of energy and momentum to count for   detector effects.  The difference is largely due 
to the lack of detector simulation in that work. Specifically, to identify at least four isolated leptons, the efficiencies (including $\eta$ cut)  are about $50\% - 60\%$ in our study,
while the parton-level analyses done in Ref.~\cite{Han:2007bk} does not  have any loss
of leptons due to lepton identification possible. In particular, the efficiencies for our analyses are lower for light invariant masses.
This is because in low mass case,  a pair of same-sign leptons
produced with significant boost could be quite difficult to discriminate from
a single lepton, as they will follow almost identical tracks and both deposit
energy in the same calorimeter cell. 
Taking this efficiency into account,  and further noting that we have not included the effects of a $K$-factor greater than 1 or other subleading methods of production for pairs of doubly charged Higgses, we find that our results are in reasonable agreement with those obtained in\,\cite{Han:2007bk}.

We note also that the $WW$ decay was studied in\,\cite{Han:2007bk} in the context of the like-sign dilepton and four jet final state, allowing a mass reconstruction for the doubly charged Higgs from the four jets. The authors conclude that, with 300 fb$^{-1}$ and assuming 100\% decay branching ratio,     three or more events are expected in this channel for Higgs masses below about 725 GeV.  Such search channel provides a nice complementary to the same-sign dilepton resonance searches for the doubly charged Higgs.

 \section{Conclusion}
\label{sec:conclusion}
In this paper, we discussed the simplified model approach to the same-sign dilepton resonance and studied its   phenomenology.  We   consider a spin 0 scalar, which  resides in a singlet, doublet or 
triplet ${\rm SU(2)}_L$ representation. The dominant pair production channel is Drell-Yan process, with 
subdominant contribution from two photon fusion process.  Doubly charged Higgs can also be singly produced via vector boson fusion, with production cross section proportional to $v^{\prime2}$, where $v^\prime$ is the vev of the neutral component.  The VBF cross section is 
usually suppressed, due to the electroweak precision constraints on $v^\prime$.     We summarized all the current direct collider search limits on the doubly charged Higgs through all six search channels.  While LEP searches reach the kinematic limit of the doubly charged Higgs mass of  around 100 GeV, the Tevatron search limits largely depend on the ${\rm SU}(2)_L$ representation that it resides in.  Only the $\mu\mu$ channel provides limits of 119 GeV, 122 GeV and 150 GeV for the singlet, the doublet and the triplet, respectively.  The other search channels, $ee$, $e\mu$, $e \tau$ and
$\mu \tau$ only provide limits for the triplet case, with no bounds obtained for the doublet and triplet cases. 

We also presented a complete set of updated indirect constraints, including Bhabha scattering, rare $\mu$ and $\tau$ decays, muonium-anti-muonium transition, and muon $g-2$.  We found that while $\mu \rightarrow eee$ provides the best limit so far, each observable probes a unique set of coupling products.  For $\mhpp=100$ GeV, products of contributing couplings are constrained to be between $10^{-3}$ to $10^{-7}$.  

We studied the LHC discovery potential of the doubly charged Higgs via Drell-Yan pair production in both the 
$eeee$ and $\mu\mu\mu\mu$ channels.  At center of mass 7 TeV with 10 ${\rm fb}^{-1}$, 3 events discovery reach is about 330 GeV, 350 GeV and 380 GeV for the singlet, the doublet and the triplet case, respectively, assuming the decay branching ratio to a given channel is 100\%.  The reach can be greatly extended to 700 GeV, 720 GeV and 800 GeV at 14 TeV center of mass energy with 100 ${\rm fb}^{-1}$ integrated luminosity.    The reach is reduced if ${\rm Br}(\hpp \rightarrow \ell \ell)$ is small due to  competing channels.

For the Higgs triplet case with non-zero $v^\prime$,  $\hpp WW$ coupling appears, which induces $\hpp \rightarrow WW$ decay, as well as the vector boson fusion production of $\hpp$.  Although the vev is constrained to be less than about 1 GeV by electroweak precision measurements, $\hpp \rightarrow WW$ competes with $\hpp \rightarrow \ell\ell$, with the corresponding decay branching ratios depending on $v^\prime$, $h_{\ell \ell^\prime}$ and $\mhpp$.  When  ${\rm Br}(\hpp \rightarrow \ell\ell)$ is small,  $\hpp \rightarrow WW$ becomes dominant, which provides a complementary discovery channel for the doubly charged Higgs. 

Note that in the simplified model approach, we decouple all other components and only keep $\hpp$ in the low energy spectrum.  If those extra states do not decouple, they will provide extra contributions to indirect measurement observables.     In cases when those states are so light that the decay of heavier $\hpp$ into light scalar states are open, ${\rm Br}(\hpp \rightarrow \ell\ell)$ will be further reduced.  The LHC study results we presented above, however, still apply.  In addition, those extra decay channels might provide novel signatures at colliders, which can be used as a complementary search to the same-sign dilepton resonance signal.

The relevant ${\rm SU(3)}_Q^J$ quantum numbers of the same-sign dilepton resonance are  ${\bf 1}_2^{0,1,2}$.  In our analyses, we studied the simplest case of a spin 0 scalar in three lowest ${\rm SU}(2)_L$ representations.  The triplet case is the one that has been studied the most in the literature.  Our analyses extended   current studies greatly by considering other two cases as well.  The simplified models that we analyzed are often the limit of a broad class of more complete models.  Our results can be applied to those models as well with minimal modifications. Higher ${\rm SU}(2)_L$ representations  could also be considered for same-sign dilepton resonance.  It is straightforward to extend our analyses to those more complicated cases as well.  Although  same-sign dilepton resonance with higher spin is also possible, it is rare to find such states in well motivated theoretical models. 

The simplified model approach provides a nice framework to accommodate and categorize experimentally measured signals, while being general enough that the results obtained in a simplified model can be applied to a broader set of models with little modification.  Once a positive experimental signal is observed, the simplified model approach 
provides a quick response to the experimental results.  It can  help us to  formulate and sharpen our understanding of the more complete theoretical models which explain the signals. In the early LHC era with rich data set available in the near future, the simplified model approach will be a powerful and useful tool to both   theoretical and experimental studies.

 \section{Acknowledgments}

We would like to thank  Tao Han and Daniel Whiteson   for useful discussions. 
VR and SS are supported by the Department of Energy
under Grant~DE-FG02-04ER-41298.  SS is also supported partly
by NSF Grants No. PHY-0653656 and 
PHYÐ0709742.

%begin{appendix}
%\section{Feynman Rules}
%\label{app:feyn}
%\end{appendix}

 %%%%%%%%%%%%%%%%%%%%%%%%%%%%%

\end{document}